\documentclass[pra,aps,twocolumn,amsfonts,showpacs,superscriptaddress,floatfix]{revtex4-1}

\usepackage{amsmath}
\usepackage{amssymb} 
\usepackage{amsfonts}
\usepackage{color} 
\usepackage{graphicx}
\usepackage{hyperref}
\usepackage{natbib}

\newcommand{\vect}[1]{\mathbf{#1}}

\begin{document}

\title{Dynamics of Nonequilibrium Dicke Models} 

\author{M. J. Bhaseen} 
\affiliation{University of Cambridge, Cavendish
  Laboratory, Cambridge, CB3 0HE, UK.}

\author{J. Mayoh} 
\altaffiliation[Present address: ]{University of
  Cambridge, Cavendish Laboratory, Cambridge, CB3 0HE, UK.}
\affiliation{School of Physics and Astronomy, University of St
  Andrews, KY16 9SS, UK.}

\author{B. D. Simons} 
\affiliation{University of Cambridge, Cavendish
  Laboratory, Cambridge, CB3 0HE, UK.}

\author{J. Keeling} 
\affiliation{School of Physics and Astronomy,
  University of St Andrews, KY16 9SS, UK.}

\date{\today}
  
\begin{abstract}
  Motivated by experiments observing self-organization of cold atoms
  in optical cavities we investigate the collective dynamics of the
  associated nonequilibrium Dicke model. The model displays a rich
  semiclassical phase diagram of long time attractors including
  distinct superradiant fixed points, bistable and multistable
  coexistence phases and regimes of persistent oscillations.  We
  explore the intrinsic timescales for reaching these asymptotic
  states and discuss the implications for finite duration experiments.
  On the basis of a semiclassical analysis of the effective Dicke
  model we find that sweep measurements over 200ms may be required in
  order to access the asymptotic regime. We briefly comment on the
  corrections that may arise due to quantum fluctuations and states
  outside of the effective two-level Dicke model description.
\end{abstract}


\maketitle

\section{Introduction}

In recent years there has been rapid progress in controlling the
behavior of ultra cold atoms using a wide variety of optical
techniques. This includes confining atoms in optical traps and optical
lattice potentials in conjunction with tremendous advances in laser
cooling. More recently it has become possible to study the properties
of Bose--Einstein condensates (BEC) in ultra high finesse optical
cavities \cite{Brennecke:Cavity}. Closely related experiments have
also been performed on novel hybrid systems combining optical fibers
on atom chips \cite{Colombe:Strong,Purdy:Integrating,Kohnen:Array}. A
central feature of these experiments is that one may access the
strongly coupled regime of cavity Quantum Electrodynamics (QED). In
this regime a large number of atoms, $N$, exchange photons many times
on the timescale set by cavity leakage. This permits the exploration
of coherent matter--light interactions and the observation of the
collective $\sqrt{N}$ splitting of the resulting eigenstates. It also
leads to novel forms of collective dynamics and cavity optomechanics
\cite{Brennecke:Opto,Ritter:Dyn,Purdy:Tunable,Brahms:Spin}.  Moreover,
the light leaving the cavity provides valuable information on strongly
correlated phases
\cite{Ritsch:Probing,Chen:Numstat,Brahms:MRI,Leslie:Trans}, thereby
fostering links between contemporary problems in cold atomic gases,
quantum optics and condensed matter physics. These systems also offer
exciting possibilities as matter--light interfaces for quantum
information processing.  This wealth of activity is further stimulated
by pioneering circuit QED experiments
\cite{Wallraff:Strong,Schuster:Resolving}, which include direct
observations of Berry' phases \cite{Leek:Berry}, vacuum fluctuations
\cite{Wallraff:Lamb}, collective behavior
\cite{Fink:Climbing,Wallraff:Collective} and three-qubit entanglement
\cite{DiCarlo:Prep}.

\begin{figure}[!t]
  \centering
  \includegraphics[width=3.2in]{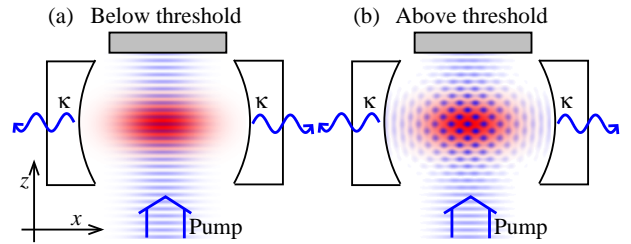}
  \caption{(color online). Experimental setup showing cold atoms (red)
    in an optical cavity with transverse pumping
    \cite{Domokos:Collective,Black:SO,Baumann:Dicke,Baumann:Dicke2}.
    (a) Below the threshold pump power only the pump mode is present.
    (b) Above the threshold the atoms self-organize into a
    checkerboard lattice and are trapped in the interference pattern
    of the pump and cavity beams. The self-organization transition for
    a BEC is described by the onset of superradiance in an effective
    nonequilibrium Dicke model.}
\label{Fig:selforg}
\end{figure}

An important aspect of these developments is the potential for novel
phases and phase transitions induced by the cavity light field.  The
latter mediates long range interactions between the atoms which may
strongly influence their behavior. It was recently predicted that an
atomic cloud with additional transverse pumping undergoes a
self-organization transition to a spatially modulated phase
\cite{Domokos:Collective}; see Fig.~\ref{Fig:selforg}.
[!t] 
This was confirmed experimentally by the Vuleti\'{c} group using
thermal clouds in an optical cavity \cite{Black:SO}. Above a critical
pumping strength the atoms self-organize to form a checkerboard
pattern as illustrated in Fig.~\ref{Fig:selforg}.  This dynamically
generated lattice leads to a strong enhancement of the cavity light
field due to coherent Bragg scattering. Heterodyne measurements of the
phase of the cavity output also reveal the discrete ${\mathbb Z}_2$
symmetry breaking of the emergent lattice.  More recently, this
self-organization phenomenon was investigated experimentally using a
BEC in an optical cavity \cite{Baumann:Dicke,Baumann:Dicke2}.  In this
setup spontaneous sublattice symmetry breaking coexists with
superfluid phase coherence giving rise to a novel form of supersolid
\cite{Andreev:SS,Chester:SS,Leggett:SS}.  In addition it was pointed
out that this self--organization transition is a dynamical version of
the superradiance transition in the Dicke model
\cite{Nagy:Dicke,Baumann:Dicke,Baumann:Dicke2}. The Dicke model
\cite{Dicke:Coherence,Hepp:Super,Wang:Dicke,Emary2003,Emary:Chaos} has
a long history and describes two-level systems or ``spins'' uniformly
coupled to light. When the matter--light coupling exceeds a critical
value the Dicke model exhibits a continuous phase transition to a
state with a non-vanishing photon population and discrete parity
symmetry breaking; for a review of the Dicke model and its
applications in quantum optics see Ref.~\cite{Garraway:Review}.

In the present cold atoms setting the effective Dicke model spin
states are two distinct momentum states of the BEC
\cite{Baumann:Dicke,Baumann:Dicke2}. Their splitting is therefore
controlled by the atomic recoil energy, and this enables the Dicke
model transition to be observed using light with optical
frequencies. This approach is a close analogue of an elegant
theoretical proposal by Dimer {\em et al} \cite{Dimer:Proposed}, for
realizing the Dicke model transition using a Raman pumping scheme
between distinct hyperfine states. These atomic experiments also
provide a direct implementation of a Dicke model Hamiltonian without
any additional diamagnetic terms. This circumvents the usual no-go
theorems for observing the superradiance transition
\cite{Birula:Nogo,Nataf:Nogo,Viehmann:SR}. The experiments also have
close connections to work on the Collective Atomic Recoil Laser (CARL)
\cite{Courtois:Recoil,Bonifacio:CARL,Bonifacio:CARL2}. For further
work on self-organized matter--light systems and the possibility of
novel phases and phase transitions in multimode cavities see
Refs.~\cite{Nagy:SO,Nagy:Self,Larson:MIRes,Larson:Stability,Gop:Emergent,Keeling:View,FV:Self,Gopal:Frus,Strack:Dicke}.

A crucial feature of the cavity superradiance experiments is that they
are intrinsically open systems with strong pumping and large cavity
loss rates \cite{Baumann:Dicke, Baumann:Dicke2}.  Any account of their
physical properties therefore requires a nonequilibrium
approach. Motivated by this experimental situation we recently
explored key aspects of the collective dynamics of BECs in optical
cavities \cite{Keeling:Collective}. On the basis of a semiclassical
analysis of the generalized Dicke model presented in
Refs.~\cite{Baumann:Dicke, Baumann:Dicke2} we obtained a surprisingly
rich phase diagram of nonequilibrium phases and phase transitions.
Most interestingly, we have found that the open system 
displays two significant features. First, for the parameters
used in the recent experiments we find additional attractors of the
long time dynamics that have {\em not} yet been seen in experiment. In
particular, the experiment suggests a normal state without photons, in
a region where the semiclassical analysis predicts that the normal
state is unstable.  Second, we find a rich array of new phases in
experimentally unexplored regions of the phase diagram.  This includes
novel coexistence phases and regimes of persistent oscillations. The
aim of this present manuscript is to shed further light on these
pertinent issues and to develop deeper links between theory and
experiment.  To this end we explore the
semiclassical collective dynamics with a specific emphasis on the
emergent timescales and the observability of the characteristic
features. A key finding is that these timescales vary quite
considerably throughout the phase diagram. Under the assumption that
the effective Dicke model fully describes the experimental system
\cite{Baumann:Dicke,Baumann:Dicke2} our primary conclusion is that the
${\mathcal O}(10{\rm ms})$ duration of the current experiments may not
be sufficient to reach the long-time asymptote in all cases. We
discuss the prospects for observing the predicted asymptotic states in
longer duration experiments and in other realizations of the
nonequilibrium Dicke model. We also discuss the possible role of
quantum fluctuations and states outside of the effective Dicke model
description.

The layout of this paper is as follows. In Sec.~\ref{Sect:Exp} we
provide an introduction to the recent experiments
\cite{Baumann:Dicke,Baumann:Dicke2} and the associated Dicke model. In
Sec.~\ref{sec:non-equil-dynam} we discuss the semiclassical dynamics
of this inherently open nonequilibrium system. In Sec.~\ref{Sect:PD}
we present the dynamical phase diagram for the presently available
experimental parameters, and we discuss the nature of the long time
attractors and the associated time evolution.  In
Sec.~\ref{sec:finiteduration} we investigate the characteristic
timescales governing the initial and final stages of the collective
dynamics, and we discuss the implications for finite duration
experiments. In Sec.~\ref{sec:genpd} we investigate the phase diagram
for a broader range of parameters and we discuss the appearance of
persistent oscillations. In Sec.~\ref{sec:discussion} we examine the
effects of contributions which go beyond the effective Dicke model and
its semiclassical treatment.  We conclude in
Sec.~\ref{sec:conclusions} and provide directions for further
research.  We also include technical appendices addressing the
derivation of the effective Dicke model, and the location of fixed
points and their linear stability properties. We also provide further
details on the phase diagram. In order to make the manuscript
self-contained we incorporate some of the principal findings of our
previous Letter \cite{Keeling:Collective}.

\section{Experiment and Nonequilibrium Generalized Dicke Model}
\label{Sect:Exp}
\begin{figure}[!t]
  \centering
  \includegraphics[width=3in]{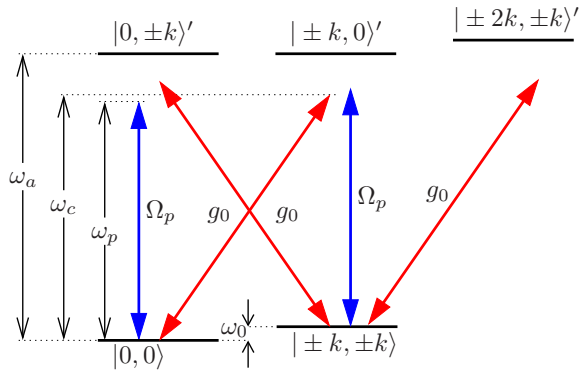}
  \caption{(color online). 
Level scheme corresponding to the
    experimental setup shown in Fig.~\ref{Fig:selforg}. The pump beam
    has frequency $\omega_p$, wavevector $k$, and Rabi frequency
    $\Omega_p$.  The strength of the cavity coupling is given by
    $g_0$.  The frequencies of the cavity and the atomic transition
    are denoted by $\omega_c$ and $\omega_a$ respectively. Here
    $|k_x,k_z\rangle$ are momentum states of the atoms in the BEC, and
    excited electronic states are denoted by a prime. The
    atomic ground state with $|k_x,k_z\rangle=|0,0\rangle$ and the
    symmetric superposition labelled as $|\!\pm k,\pm k\rangle$
    constitute an effective two-level system governed by an effective
    nonequilibrium Dicke model. The corresponding level splitting is
    given by $\omega_0=2\omega_r$, where $\omega_r=\hbar^2k^2/2m$ is
    the atomic recoil energy resulting from the absorption or emission
    of a single photon. In general, multi-photon processes are
    required in order to couple the ground state to higher momentum
    states.}
  \label{fig:level-scheme}
\end{figure}

The experiment of Ref.~\cite{Baumann:Dicke} consists of a $^{87}$Rb
BEC with approximately $N=10^5$ atoms prepared in their motional
ground state with $|k_x,k_z\rangle=|0,0\rangle$. The atoms are placed
in an ultra high finesse optical cavity of length $178\,\mu{\rm m}$
and cavity loss rate $\kappa=8.1$\,MHz.  As shown in
Fig.~\ref{Fig:selforg} the BEC is subjected to a transverse pump beam
with Rabi frequency $\Omega_p$, wavevector $k$ and frequency
$\omega_p$. The latter is far detuned from the atomic transition
frequency $\omega_a$, in order to avoid population of this excited
level. One may therefore neglect the effects of spontaneous emission.
However, the pump frequency is near detuned to the cavity frequency
$\omega_c$, resulting in efficient scattering from the pump beam into
the cavity and vice versa. The coupling strength of a single atom to
the cavity mode is denoted by $g_0$ and the corresponding level scheme
is shown in Fig.~\ref{fig:level-scheme}.
The experiment is a close analogue of a theoretical proposal by Dimer
{\em et al} \cite{Dimer:Proposed} for realizing the Dicke model by
using Raman pumping to couple to two ground state hyperfine levels.  A
notable difference is that the present experiment exploits a Rayleigh
scheme, involving distinct momentum states rather than internal
hyperfine states. This generically leads to the
presence of a back-reaction term, discussed below, which may be
avoided in the proposal of Ref.~\cite{Dimer:Proposed}.

Absorption and emission of photons yields an effective two-level
``spin'' system \cite{Baumann:Dicke,Nagy:Dicke} where spin down
corresponds to the ground state $|0,0\rangle$, and spin up corresponds
to the excited momentum state $|\!\pm k,\pm k\rangle\equiv
\frac{1}{2}\sum_{\alpha,\beta=\pm}|\alpha k,\beta k\rangle$.  The
latter denotes the symmetric superposition of momentum states
resulting from two-photon emission and absorption processes with
$k_x,k_z\in \{\pm k\}$. In this basis one may introduce collective
spin raising operators $S^+=\sum_{i=1}^N|\pm k,\pm k\rangle_i \,
_i\langle 0,0|$ where $i$ labels the atoms and $S^-$ is obtained by
Hermitian conjugation.  The quantum dynamics of this inherently open
system, with a large cavity loss rate $\kappa$, can be described by
the density matrix equation in Lindblad form \cite{Scully:QO}
\begin{equation}
    \label{eq:1}
    \partial_t \rho = - i [H, \rho] - \kappa \left(
      \psi^\dagger \psi \rho - 2 \psi \rho \psi^\dagger + \rho \psi^\dagger \psi 
    \right),
\end{equation}
where $\rho$ is the system density matrix, $\psi$ is the cavity photon
mode annihilation operator, and $t$ denotes time
\footnote{In these notations the rate of loss of energy is $2\kappa$.}. 
The effective Hamiltonian, $H$ takes the form of a
generalized Dicke model with
\cite{Baumann:Dicke,Nagy:Dicke,Dimer:Proposed}
\begin{multline}
  H = \omega \psi^\dagger \psi + \omega_0S_z + US_z  \psi^\dagger \psi 
  + g
  (\psi^\dagger S^- + \psi S^+)
  \\+ g^\prime
  (\psi^\dagger S^+ + \psi S^-),
  \label{Dickeham}
\end{multline}
where ${\bf S}=(S_x,S_y,S_z)$ is the effective collective spin of
length $N/2$ and $S^\pm=S_x\pm iS_y$. The derivation of
Eq.~(\ref{Dickeham}) from the microscopic description, together with a
discussion of higher order contributions, is given in
Appendix~\ref{sec:depart-from-gener}.  For weak pumping, and in the
limit that the atom-cavity detuning is much larger than both the
pump-cavity detuning and the recoil energy
\cite{Baumann:Dicke,Baumann:Dicke2}, the coefficients are given by
$\omega=\omega_c - \omega_p - N (5/8) g_0^2/(\omega_a-\omega_c)$ and
$\omega_0=2\omega_r$, where $\omega_r=\hbar^2k^2/2m$ is the atomic
recoil energy.  The term involving $U=-(1/4)
g_0^2/(\omega_a-\omega_c)$ describes the back-reaction of the cavity
light field on the BEC, and may be interpreted as the AC Stark shift
due to the appearance of a weak optical lattice in the cavity.  In the
experiment \cite{Baumann:Dicke} both the pump and the cavity are red
detuned from the atomic transition, so $U$ is negative. However, both
signs of $U$ are physically achievable. In the atomic ground state,
the effective cavity frequency $\omega_{\rm eff}=\omega+US_z$ is given
by $\omega_{\rm eff}=\omega
-UN/2=\omega_c-\omega_p-Ng_0^2/2(\omega_a-\omega_c)$ \footnote{In the
  notations of Ref.~\cite{Baumann:Dicke} the effective cavity
  frequency $\omega_{\rm eff}$ is denoted by $\omega$.}, in agreement
with Ref.~\cite{Baumann:Dicke}.  The Hamiltonian (\ref{Dickeham})
contains both co- and counter-rotating matter--light couplings denoted
by $g$ and $g^\prime$ respectively.  In the large atom--pump detuning
limit relevant to the experiment \cite{Baumann:Dicke} one may write
$g=g^\prime=g_0\Omega_p/2(\omega_p-\omega_a)$.

For the experimental parameters used in Ref.~\cite{Baumann:Dicke},
$\omega_0=0.047$MHz and $UN=-6.5\kappa/4=-13.3$MHz, where the latter
is inferred from the observed dispersive shift of the cavity
frequency, $\omega_{\rm eff}=\omega_c-\omega_p+2UN$ \footnote{In the
  experiment \cite{Baumann:Dicke} the spatial overlap of the cavity
  mode profile and atomic density is not perfect, and so this feedback
  term is reduced slightly from this ideal value.}.  In the subsequent
discussion we will endevor to place these experiments in a broader
context, and without loss of generality we will approximate these
conditions as $\omega_0\approx 0.05$MHz and $UN\approx -10$MHz; note
that the latter differs from the value taken in our previous Letter
\cite{Keeling:Collective} due to a small discrepancy in the reported
Hamiltonian in Ref.~\cite{Baumann:Dicke}. Specifically, the
Hamiltonian given in Eq.~(\ref{Dickeham}) differs from Eq.~(4) of
Ref.~\cite{Baumann:Dicke} due to a discrepancy in the indicated matrix
element $M=3/4$ in the notation of Ref.~\cite{Baumann:Dicke}.  This
does not affect the location of the reported superradiance transition,
but is important for establishing the broader phase diagram. In
addition to the energy and timescales appearing in the model described
by Eqs.~(\ref{eq:1}) and (\ref{Dickeham}), there is a limit on the
duration of current experiments which is set by the rate of atom
loss. In the initial experiments \cite{Baumann:Dicke} this was of the
order of $100$ms, but is notably longer in subsequent experiments
\cite{Baumann:Dicke2}.

\section{Semiclassical Dynamics of The Open System}
\label{sec:non-equil-dynam}

Having discussed the effective Hamiltonian and the density matrix
equation of motion in Sec.~\ref{Sect:Exp}, we now turn to discuss the
dynamics arising from this model. This is essential in order to
interpret time dependent nonequilibrium experiments performed in an
open cavity \cite{Baumann:Dicke,Baumann:Dicke2}. In view of the large
number of atoms comprising the Dicke model spin states, we will first
consider the semiclassical limit of this dynamics
\cite{Keeling:Collective}.  In Sec.~\ref{sec:discussion} we will
briefly comment on the role of quantum fluctuations.

\subsection{Equations of Motion and Symmetries}
\label{Sect:Dynam}
The semiclassical equations of motion for the open system described by
Eqs.~(\ref{eq:1}) and (\ref{Dickeham}) are given by
\begin{equation}
\begin{aligned}
  \dot{S}^- &= 
  - i (\omega_0 + U |\psi|^2 ) S^-
  + 2 i (g \psi + g^\prime \psi^\ast) S_z,
  \\
  \dot{S}_z &=
  - i g \psi S^+ + i g \psi^\ast S^-
  + i g^\prime \psi S^- - i g^\prime \psi^\ast S^+,
  \\
  \dot{\psi} &= 
  - \left[ \kappa + i (\omega + U S_z) \right] \psi
  - i g S^- - i g^\prime S^+,
\end{aligned}
\label{eqmo}
\end{equation} 
where $S^\pm\equiv S_x\pm iS_y$, $\kappa$ is the cavity loss rate, and
we neglect the effects of atom loss \cite{Baumann:Dicke}. The
Hamiltonian in Eq.~(\ref{Dickeham}) conserves the total length of the
collective spin since $[{\bf S}^2,S_\alpha]=0$ for $\alpha=x,y,z$.
Likewise,  Eq.~(\ref{eqmo}) satisfies
$\partial_t {\bf S}^2=0$ for all $\kappa$. As such, the dynamics can
be explored on the Bloch sphere with $|{\bf S}|=N/2$. In addition to
this conservation law there are further discrete symmetries. In
particular, the equations of motion in Eq.~(\ref{eqmo}) are invariant
under the parity transformation
\begin{equation}
\psi\rightarrow -\psi,\quad S^\pm\rightarrow -S^\pm,
\label{parity} 
\end{equation}
as in the equilibrium Dicke model. This symmetry is
spontaneously broken on passing from the normal phase with $\psi=0$ to
the superradiant phase with $\psi\neq 0$
\cite{Baumann:Dicke,Baumann:Dicke2}.   The equations of
motion are also invariant under the combined variable and parameter change
\begin{equation}
{\bf S}\rightarrow -{\bf S}, \quad \psi\leftrightarrow \psi^\ast, \quad 
\omega\rightarrow-\omega, \quad g\leftrightarrow g^\prime.
\label{duality} 
\end{equation}
As we shall see in Sec.~\ref{sec:steady-states}, both the symmetry in
Eq.~(\ref{parity}) and the duality relation in Eq.~(\ref{duality})
will have a direct manifestation on the Bloch sphere portraits; the 
attractors are related by these discrete transformations.

In order to get a full understanding of the behavior of
Eq.~(\ref{eqmo}), it is necessary to address two questions.  The first
regards the nature of the long time attractors.  The second concerns
the full time evolution and its connection to the asymptotics. The
former is important because there are fundamental differences between
the dynamical phase diagram of the open system and the equilibrium
phase diagram of the Hamiltonian; even for $\kappa\rightarrow 0$ these
are generically distinct.  The latter question is crucial because the
open cavity experiments have a finite duration and may not always
reach the long time asymptote.

\begin{figure}[!t]
\begin{center}
\includegraphics[width=3.2in,clip]{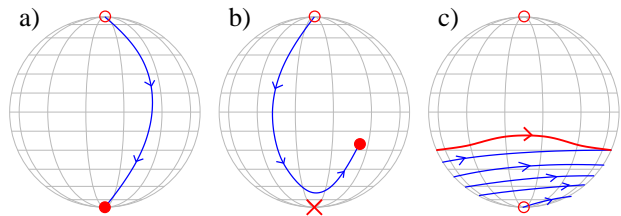}
\caption{(color online).  
Schematic illustration of the different
  types of behavior displayed by the semiclassical equations of motion
  in Eq.~(\ref{eqmo}), for trajectories on the Bloch
  sphere with $|{\bf S}|=N/2$.  (a) Evolution from an unstable fixed point
  (open circle) to a stable attractor (closed circle) of the long time
  dynamics for all initial conditions. Both the stable and unstable
  points have $\dot{\bf S}=0$ and $\dot\psi=0$. However, for
  the stable attractor small perturbations decay, while for the
  unstable fixed point fluctuations grow.  (b) As for (a) but with a
  hyperbolic fixed point (cross) having one stable and one unstable
  eigenmode. Paths first approach and then leave the vicinity of
  this hyperbolic point, before eventually reaching a stable attractor
  with $\dot{\bf S}=0$ and $\dot\psi=0$.  (c)
  Dynamics exhibiting a stable limit cycle with $\dot{\bf S}\neq
  0$ and $\dot\psi\neq 0$ for all initial conditions.}
\label{Fig:fixedpointvisual}
\end{center}
\end{figure}

For most values of the parameters, the long-time asymptotes are steady
states and may be identified as stable fixed points. That is to say,
for these parameters there are values of ${\bf S}$ and $\psi$ for
which $\dot{\bf S}=0$ and $\dot\psi=0$, and these steady states are
the eventual fate of the semiclassical dynamics for all initial
conditions.  We will therefore discuss what these steady states are,
and where possible, give analytical formulae for them.  In the
following sections we will then use this information to present
dynamical phase diagrams, describing which stable fixed points exist
for different values of the parameters.  We will then
address the dynamical bifurcations that correspond to the phase
boundaries, as well as addressing those cases where the long time
asymptotes are more complicated than steady states, such as persistent
oscillations. We illustrate these possibilities in
Fig.~\ref{Fig:fixedpointvisual}.

Before discussing the semiclassical
equations of motion in Eq.~(\ref{eqmo}), let us  comment on
some known limiting cases that have been studied in the
literature. With $\kappa=0$, $g^\prime=0$ and $U=0$, these equations
correspond to the semiclassical dynamics of the equilibrium Dicke
model without the loss of photons and without counter-rotating terms
\cite{Bonifacio:Coherent}. This model arises in various contexts and
has recently been discussed in relation to nonequilibrium Cooper
pairing
\cite{Andreev:Noneqfesh,Barankov:Coll,Yuzbashyan:Solution,Yuzbashyan:FB}.
In this setting, the constituent fermions are modelled by Anderson
pseudo spins \cite{Anderson:RPA}, and the cavity light field $\psi$
corresponds to the closed molecular channel.  A key finding of these
studies is the presence of collective oscillations.  This may also be seen by
exploiting the integrability of the closely related BCS
(Bardeen--Cooper--Schrieffer) Hamiltonian
\cite{Yuzbashyan:Solution,Yuzbashyan:FB}.  The same equations of
motion also apply to polariton condensation and the synchronisation of
oscillators \cite{Eastham:New,Eastham:Phase}. Quantum corrections to
this collective dynamics have also been explored in
Refs.~\cite{Keeling:QCorr,Babelon:Semi}.

In contrast to the case when $g^\prime=0$, when $\kappa=0$,
$g=g^\prime$ and $U=0$ the equilibrium Dicke model is no longer
integrable. Nonetheless, the model is tractable in the thermodynamic
limit and displays a mean field superradiance transition.  Strikingly,
the energy levels reveal a crossover from Poisson statistics to Wigner
statistics in the vicinity of the critical coupling
\cite{Emary2003,Emary:Chaos}. This indicates the onset of quantum
chaos, and is accompanied by chaotic attractors in the analogous
classical dynamics.

More recently, Dimer {\em et al} \cite{Dimer:Proposed} have proposed a
novel scheme for realizing the nonequilibrium Dicke model described
by Eqs.~(\ref{eq:1}) and (\ref{Dickeham}) with $\kappa\neq 0$. The
parameters in this effective model are readily adjustable and they
focus on the particular case with $g=g^\prime$ and $U=0$. A notable
observation is that cavity losses lead to a shift of the mean field
superradiance transition, in agreement with recent experiments with
$U\neq 0$ \cite{Baumann:Dicke,Baumann:Dicke2,Nagy:Dicke}.

It is evident from this survey of limiting cases that rich
collective dynamics is expected to emerge for the more general system
of equations given by Eq.~(\ref{eqmo}), and in open cavity
experiments.  We shed light on this below.

\subsection{Fixed Point Attractors}
\label{sec:steady-states}

In order to get a handle on the possible long time steady states, we
first enumerate the solutions of the equations of motion with
$\dot{\bf S}=0$ and $\dot\psi=0$. These fixed point solutions may
either be stable or unstable, and we postpone a discussion of their
stability properties until Sec.~\ref{sec:stability}.  
It is readily verified that the normal state ($\Downarrow$) is always
a possible steady state solution with all the spins pointing down,
$S_z=-N/2$ and no photons, $\psi=0$.  Likewise, so is the inverted
state ($\Uparrow$) with all the spins pointing up, $S_z=N/2$ and no
photons, $\psi=0$. More generally one may look for non-trivial
solutions with a non-vanishing photon population and a non-trivial
magnetization, $S_z$. To find these configurations we first note that
a steady state solution satisfying the first equation in Eq.~(\ref{eqmo}) automatically satisfies the
second equation. As such Eq.~(\ref{eqmo}) reduces to a pair of complex
equations. Denoting $\psi=\psi_1+i\psi_2$ and $S^\pm=S_x\pm iS_y$ one
obtains
\begin{equation}
\begin{aligned}
\left[\omega_0 +U(\psi_1^2+\psi_2^2)\right] & (S_x-iS_y)  = \\
& 2\left[(g+g^\prime)\psi_1+i(g-g^\prime)\psi_2\right]S_z,
\label{eqmo1}
\end{aligned}
\end{equation}
and
\begin{equation}
[\kappa+i(\omega+US_z)](\psi_1+i\psi_2) 
=-i(g+g^\prime)S_x-(g-g^\prime)S_y.
\label{eqmo2}
\end{equation}
In general these equations may be difficult to solve
analytically. However simplifications occur when $U=0$ or when
$g=g^\prime$. We focus here on the latter since the experiments of
Ref.~\cite{Baumann:Dicke} correspond to $g=g^\prime$ and negative $U$.
We discuss the behavior for $g\neq g^\prime$ in Appendix
\ref{App:Genfp}.

With $g=g^\prime$ the fixed point Eqs. (\ref{eqmo1}) and (\ref{eqmo2}) read
\begin{subequations}
\begin{align}
(\omega_0+U|\psi|^2)S_x & = 2g(\psi+\psi^\ast)S_z, \label{ssggua}\\
(\omega_0+U|\psi|^2)S_y & = 0, \label {ssggub}\\
(\omega+US_z-i\kappa)\psi & =-2gS_x. \label{ssgguc}
\end{align}
\label{ssggu}
\end{subequations}
It follows from Eq.~(\ref{ssggub}) that there are two classes of 
solutions depending on whether $S_y=0$ or
$\omega_0+U|\psi|^2=0$.  We consider these two classes in
turn and refer to the non-trivial steady state solutions as
superradiant A (SRA) and  B (SRB) respectively.  Assuming
$\omega_0>0$, for $U\ge 0$ only the first type of SRA solution may be
present. This solution corresponds to the familiar superradiant phase
in the usual Dicke model where $U=0$. For $U<0$ the second type of SRB
solution may exist \cite{Keeling:Collective}. As we shall discuss
below, in general these solutions are continuously connected in the
broader parameter space with $g\neq g^\prime$.  Nonetheless, it is
important to distinguish between these distinct solutions of the
steady state equations of motion when $g=g^\prime$. We will discuss
the experimental consequences of this distinction in
Secs.~\ref{Sect:PD} and \ref{sec:finiteduration}.
\subsubsection{Superradiant A (SRA) Steady States}
Equation (\ref{ssgguc}) may be rearranged as an equation for
$\psi$ and substituted into Eq.~(\ref{ssggua}):
\begin{equation}
\omega_0(\omega+US_z)^2+\omega_0\kappa^2+4g^2US_x^2=-8g^2(\omega+US_z)S_z,
\label{sraquad}
\end{equation}
where we have cancelled a factor of $S_x\neq 0$ from both sides of the
equation. Using the fixed length spin constraint, $S_x^2=N^2/4-S_z^2$
one obtains a quadratic equation for $S_z$.  This may be solved to
yield \cite{Keeling:Collective}
\begin{equation}
\frac{S_z}{N}=-\frac{\omega}{UN}\pm\sqrt{\frac{g^2N[4\omega^2-(UN)^2]-\omega_0UN\kappa^2}{(UN)^2(\omega_0UN+4g^2N)}},
\label{szsra}
\end{equation}
where the accompanying steady state photon population follows from
Eq.~(\ref{ssgguc}). In general only one of these roots corresponds to
a physical solution with $|{\bf S}|\le N/2$. However, as we shall
discuss in Sec.~\ref{sec:nature-attractors-u0} and
Appendix~\ref{App:Bifurcations}, there are regions of parameter space
where both roots of Eq.~(\ref{szsra}) are supported; see the regions
denoted 2SRA in Figs.~\ref{srasrbtran} and \ref{trizoom}. In addition,
there are two possible signs for $S_x=\pm \sqrt{N^2/4-S_z^2}$, where
the associated sign of $\psi$ is determined by
Eq.~(\ref{ssgguc}). This sign choice corresponds to the parity symmetry in
Eq.~(\ref{parity}) which is spontaneously broken at the superradiance
transition.

The critical coupling strength corresponding to the onset of
superradiance starting from the normal ($\Downarrow$) or inverted
state ($\Uparrow$) is obtained by setting ${\bf S}=(0,0,\mp N/2)$ in
Eq.~(\ref{sraquad}). One obtains
\begin{equation}
g_a^\mp\sqrt{N}=\sqrt{\pm\frac{\omega_0(\omega_\mp^2+\kappa^2)}{4\omega_\mp}},
\label{sragc}
\end{equation}
where $\omega_\mp\equiv \omega\mp \omega_u$ and $\omega_u\equiv
UN/2$. It is readily seen from the Hamiltonian (\ref{Dickeham}) that
$\omega_\mp$ plays the role of an effective cavity frequency for the
normal and inverted states respectively. For the special case where
$U=0$ this agrees with the results of Dimer {\em et al}
\cite{Dimer:Proposed}. In the additional limit $\kappa=0$,
Eq.~(\ref{sragc}) reproduces the location of the superradiance
transition, $g\sqrt{N}=\sqrt{\omega\omega_0}/2$ for the equilibrium
Dicke model with counter-rotating terms. More generally,
Eq.~(\ref{sragc}) gives the onset of the SRA phase in the open cavity
system with transverse pumping and $g=g^\prime$, as recently confirmed
experimentally \cite{Baumann:Dicke}. The explicit dependence on
$\kappa$ of the phase boundary in Eq.~(\ref{sragc}) emphasizes the
open character of the experimental system.

\subsubsection{Superradiant B (SRB) Steady States}
\label{sec:srb}
For negative $U$ it is evident from Eq.~(\ref{ssggub}) that another
class of solutions may be obtained if $\omega_0+U|\psi|^2=0$. Equation
(\ref{ssggua}) may thus be fulfilled by taking $\psi$ to be purely
imaginary. It then follows from Eq.~(\ref{ssgguc}) that
$\omega+US_z=0$. This yields \cite{Keeling:Collective}
\begin{equation}
\psi=\pm i\sqrt{-\frac{\omega_0}{U}},\quad S_z=-\frac{\omega}{U}, \quad 
S_x=\mp\frac{\kappa}{2g}\sqrt{-\frac{\omega_0}{U}},
\label{srbsol}
\end{equation}
where the magnitude of $S_y$ follows from the normalization
condition ${\bf S}^2=N^2/4$. In order to obtain real solutions for $S_y$, 
we require $S_x^2+S_z^2\le N^2/4$. This is equivalent to the condition 
$g\ge g_b$ where
\begin{equation}
g_b\sqrt{N}=\kappa\sqrt{\frac{\omega_0\omega_u}{2\left(\omega^2-\omega_u^2\right)}}.
\label{gb}
\end{equation}
In order to yield $|S_z|<N/2$ we require
$|\omega|<|\omega_u|$.  In contrast to the SRA solution which may
exist for either sign of $U$ depending on the parameters, the
functional dependence in Eq.~(\ref{srbsol}) clearly indicates that the
SRB solution only exists for $U<0$. In the special case where $g=g_b$
and $S_y=0$, the SRA and SRB solutions coincide.

In conjunction with both possible signs for $S_y=\pm |S_y|$,
Eq.~(\ref{srbsol}) defines four distinct steady states. These divide
into two pairs of solutions, where the pairs of solutions are related
by the discrete parity symmetry in Eq.~(\ref{parity}). As we shall see
in Sec.~\ref{sec:nature-attractors-u0}, two of these four solutions
correspond to stable attractors of the long time dynamics whilst the
other two solutions correspond to unstable fixed points; see
Fig.~\ref{fig:characteristic-trajectories}(d).

\subsection{Linear Stability of Fixed Points and More Exotic Attractors}
\label{sec:stability}

In Sec.~\ref{sec:steady-states} we discussed the possible fixed points
of the equations of motion with $\dot{\bf S}=0$ and $\dot\psi=0$. Here
we turn our attention to the linear stability of these fixed points as
potential candidates for the long time attractors.  The calculations
are most transparent if we consider the instability of the normal
($\Downarrow$) and inverted states ($\Uparrow$) where $\psi=0$ and
$S_z=\mp N/2$ respectively. For arbitrary fixed points the approach is
readily generalized but is algebraically more involved; the details
are outlined in Appendix \ref{App:linstab}.  Writing
$\psi=\psi_0+\delta\psi$ and $S^-=S_0^-+\delta S^-$ where $\psi_0=0$,
$S_0^-=0$, $S_z=\mp N/2$, and substituting into Eq.~(\ref{eqmo}) one
obtains the linearized equations
\begin{equation}
\begin{aligned}
\dot{\delta S^-} & =  -i\omega_0\delta S^-\mp igN\delta\psi\mp ig^\prime N\delta\psi^\ast,\\
\dot{\delta\psi} & =  -(\kappa+i\omega_\mp)\delta\psi-ig\delta S^--ig^\prime \delta S^+,
\end{aligned}
\end{equation}
where ${\dot S}_z=0$, $\omega_\mp\equiv \omega\mp \omega_u$ and
$\omega_u\equiv UN/2$. Parameterizing $\delta\psi=a e^{-i\eta t}+b^\ast
e^{i\eta^\ast t}$ and $\delta S^-=c e^{-i\eta t}+ d^\ast e^{i\eta^\ast
  t}$ and equating coefficients with the same time dependence one
obtains algebraic equations for $a$, $b$, $c$ and $d$. The 
self-consistency equations characterize the
possible instabilities  and are given by
Eqs.~(\ref{niselfcon}) and (\ref{linquart}). In
the case where $g=g^\prime$ the frequencies $\eta$ satisfy
\begin{equation}
  (\eta^2-\omega_\mp^2-\kappa^2)(\eta^2-\omega_0^2)\mp 4g^2N\omega_\mp\omega_0
  -2i\kappa\eta(\eta^2-\omega_0^2)=0.
\label{equalgquart}
\end{equation} 
The dividing line between exponentially growing and decaying
fluctuations corresponds to Eq.~(\ref{equalgquart}) having real
solutions for $\eta$. In this case the imaginary part of
Eq.~(\ref{equalgquart}) vanishes when either $\eta=0$ or
$\eta^2=\omega_0^2$. Demanding that the real part of
Eq.~(\ref{equalgquart}) vanishes when $\eta=0$ yields
Eq.~(\ref{sragc}). That is to say, the normal and inverted states
become unstable at precisely the same point as the SRA state becomes
possible. For $g\ge g_a^\mp$, Eq.~(\ref{equalgquart}) has one unstable
root and in the language of dynamical phase transitions this
corresponds to a pitchfork bifurcation.  In the case of frequencies
satisfying $\eta^2=\omega_0^2$ the real part of
Eq.~(\ref{equalgquart}) vanishes when $\omega_\mp=0$. This implies
that the normal and inverted states also become unstable when
$\omega=\pm UN/2$ respectively. For values of $\omega$ beyond these
points Eq.~(\ref{equalgquart}) develops two unstable roots and the
dynamical phase transition corresponds to a Hopf bifurcation.  As we
shall see in Secs.~\ref{Sect:PD} and \ref{sec:genpd}, all of these
instabilities describe boundaries in the emergent dynamical phase
diagrams shown in Figs.~\ref{fig:omega-g-negative-u} and
\ref{fig:omega-g-positive-u}.

In the above analysis we have outlined the existence of various fixed
points and briefly discussed their linear stability properties. These
considerations are essential because more than one of these fixed
points may exist at a given point in parameter space. For example, the
normal ($\Downarrow$) and inverted ($\Uparrow$) fixed points always
exist, possibly as unstable fixed points, even in the presence of the
superradiant solutions.  As we shall see in Sec.~\ref{Sect:PD}, there
are in fact cases where more than one stable fixed point exists at a
given point in parameter space.  
In addition to these coexistence phases, where the final state depends on
the initial conditions, it is also possible to find regions of
parameter space where no stable fixed point exists. In these cases the
system may be attracted to time dependent solutions such as limit
cycles, as found in other nonlinear dynamical systems.  In the
remainder of this manuscript we search for the complete set of stable
attractors of the long time dynamics including fixed points, bistable
and multistable coexistence phases, and time dependent trajectories.

\section{Dynamical Phase Diagram of Long Time Attractors for
  $g=g^\prime$ and $U<0$}
\label{Sect:PD}

In the previous section we gave a brief overview of the simplest fixed
point attractors and their linear stability properties. In this
section we build upon these results and establish the dynamical phase
diagram corresponding to the semiclassical dynamics in
Eq.~(\ref{eqmo}). That is to say, we identify which stable long time
attractors exist at a given point in parameter space.  In order to
make contact with experiment \cite{Baumann:Dicke,Baumann:Dicke2} in
this section we restrict our attention to $g=g^\prime$ and $U<0$. We
begin in Sec.~\ref{sec:phase-diagr-exper} by exploring the phase
diagram as a function of the remaining parameters, $g, \omega$ and
$U$, where the value of $\omega_0=0.05$MHz is motivated by
Ref.~\cite{Baumann:Dicke}.  In Sec.~\ref{sec:nature-attractors-u0} we
then focus on different points in the dynamical phase diagram in order
to clearly expose the nature of the underlying attractors, including
their stability properties and their locations on the Bloch sphere.
In Sec.~\ref{sec:timeevol} we then discuss the characteristic time
evolution towards the stable asymptotic states, and we rationalize
these findings using linear stability analysis. In some regimes of
parameter space, the time evolution can be rather slow, and we
characterize where this occurs. We discuss the significant
implications of these regions of long-lived transients for finite
duration experiments in Sec.~\ref{sec:finiteduration}.

\subsection{Phase Diagram of Asymptotic Stable Attractors}
\label{sec:phase-diagr-exper}

\begin{figure}[!t]
  \centering
  \includegraphics[width=3.2in]{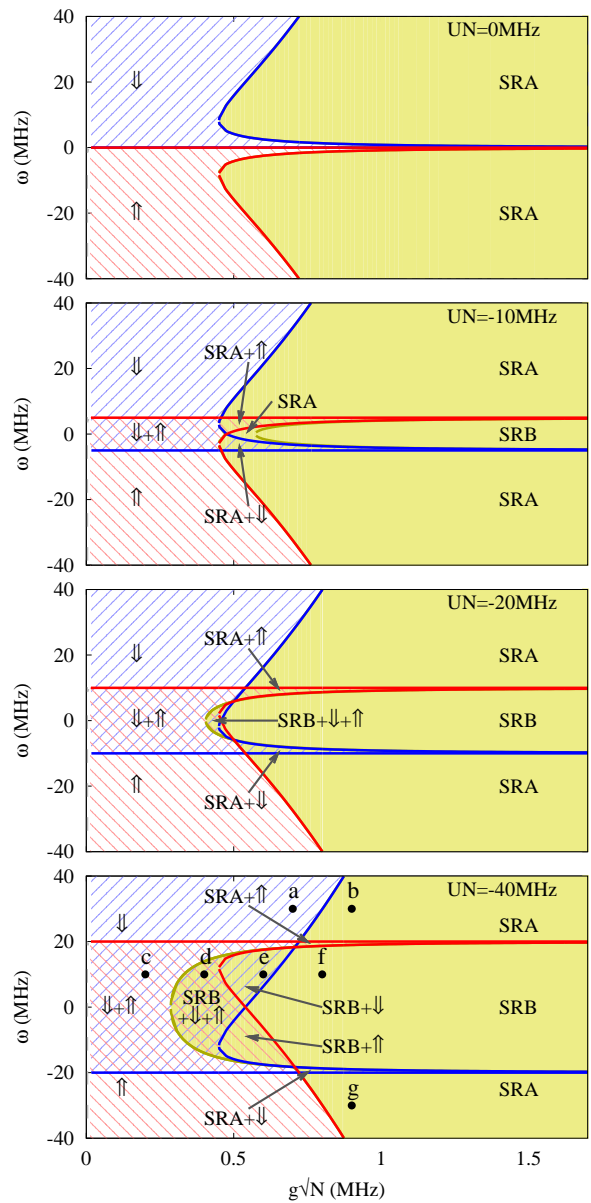}
  \caption{(color online). 
Dynamical phase diagram of the stable
    attractors as a function of the cavity frequency $\omega$ and the
    coupling $g=g^\prime$ with parameters $\kappa=8.1${\rm
      MHz} and $\omega_0=0.05$MHz taken from
    Ref.~\cite{Baumann:Dicke}.  The panels represent different values
    of the feedback term $U$, going from $U=0$ (top) to $UN=-40$MHz
    (bottom). The second panel corresponds to the experimental
    parameters used in Ref.~\cite{Baumann:Dicke}.  Points (a) - (f)
    marked in the bottom panel correspond to the fixed point
    illustrations shown in Fig.~\ref{fig:characteristic-trajectories}.
    The characteristic time evolution at points (b), (f) and (g) is
    given in Figs.~\ref{fig:time-evolution} and
    \ref{fig:time-evo-long-wait}.}
  \label{fig:omega-g-negative-u}
\end{figure}

As discussed in Sec.~\ref{Sect:Exp} the experiments of
Ref.~\cite{Baumann:Dicke} are performed with $\omega_0\simeq 0.05$MHz,
$\kappa=8.1$MHz, $UN\simeq -10$MHz and $g=g^\prime$.  We therefore
summarise the dynamical phase diagram with these parameters.  In order
to provide some orientation, we illustrate how this phase diagram
relates to that for the open Dicke model with $U=0$, as a function of
$\omega$ and $g$.  We consider fixed values of the feedback term $U$,
starting from the simplest possible case with $U=0$, and decrease this
parameter through the experimental value; see
Fig.~\ref{fig:omega-g-negative-u}.

In this nonequilibrium setting, each phase is labelled according to
the complete set of stable long time attractors of the semiclassical
dynamics given in Eq.~(\ref{eqmo}). That is to say, the phase diagram
corresponds to starting the system in a wide variety of initial
conditions and examining the totality of stable end points.
In this respect, the phase boundaries should be thought of as
dynamical phase transitions, which separate distinct regimes of
asymptotic behavior. In particular, the blue and red boundaries in
Fig.~\ref{fig:omega-g-negative-u} correspond to the instability of the
normal ($\Downarrow$) and inverted ($\Uparrow$) states respectively,
and are given by Eq.~(\ref{sragc}), whilst the accompanying horizontal
segments correspond to $\omega_\mp=0$. The gold phase boundary
indicates the critical coupling for the onset of SRB and is given by
Eq.~(\ref{gb}).  It is important to emphasize that whilst all of these
dynamical phase boundaries may be investigated experimentally, not all
of them will emerge in a given experiment; the relevant phase
boundaries are determined by the initial conditions. In particular,
the experiments performed so far all begin in the normal state
($\Downarrow$) with no photons \cite{Baumann:Dicke,Baumann:Dicke2}.
However, this is not a fundamental experimental restriction, and it is
essential to survey the totality of attractors for all initial
conditions, before considering particular initial states.

For $U=0$ and $\omega >0$, the structure of
Fig.~\ref{fig:omega-g-negative-u} mirrors the equilibrium phase
diagram of the Dicke model, having a transition from a phase at low
$g$ where only the normal state ($\Downarrow$) is possible, to a phase
where only the SRA state occurs.  In the terminology of dynamical
systems this particular dynamical phase transition occurs via a
pitchfork bifurcation at
$g\sqrt{N}=\sqrt{\omega_0(\omega^2+\kappa^2)/4\omega}$
\cite{Dimer:Proposed}; a pair of superradiant fixed points emerge when
the normal state loses stability. This parallels the situation in the
equilibrium Dicke model where a pair of parity related superradiant
solutions emerge at a continuous phase transition. It is notable that
as $\omega \to 0$, the critical value of $g$ required for
superradiance tends to infinity. This is because for $g=g^\prime$ only
the real part of $\psi$ drives the polarization of the two-level
system via the collective coupling $g(\psi+\psi^\dagger)(S^++S^-)$; as
$\omega \to 0$, $\psi$ becomes purely imaginary as may be seen from
Eq.~(\ref{eqmo}). In addition, for $\omega<0$, the open dynamical
system shows behavior that could not occur in thermal equilibrium; the
normal state ($\Downarrow$) becomes unstable and the inverted state
($\Uparrow$) with $S_z = N/2$ and $\psi=0$ is stable instead.  It is
evident from the Hamiltonian in Eq.~(\ref{Dickeham}) that the inverted
state is of higher energy than the normal state.  However, in contrast
to the suggestions of Ref.~\cite{Liu:Lightshift}, the relevant
question for the experimental realization is dynamical stability, as
opposed to minimum free energy. This behavior is directly confirmed by
the duality of the equations of motion given in Eq.~(\ref{duality}),
and is reflected in the $\omega\rightarrow -\omega$ inversion symmetry
of Fig.~\ref{fig:omega-g-negative-u}.

\begin{figure}[!t]
 \centering
 \includegraphics[width=3.2in]{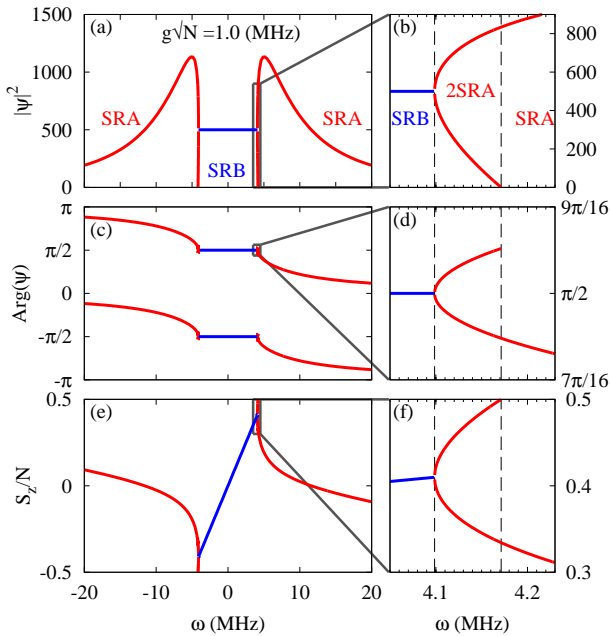}
 \caption{(color online).  
Vertical slice through the second panel of
   Fig.~\ref{fig:omega-g-negative-u} with $UN=-10$MHz, corresponding to
   the experimental parameters used in Ref.~\cite{Baumann:Dicke}. 
Variation of (a) $|\psi|^2$, (c) ${\rm Arg}(\psi)$ and (e) $S_z$ on
   passing from the SRA to SRB phases. In the vicinity of both of 
these transitions there is a narrow region of bistability
   denoted as 2SRA, where both of the SRA solutions given in
   Eq.~(\ref{szsra}) coexist; see panels (b), (d) and (f) for
   magnified images  of the highlighted regions.}
\label{srasrbtran}
\end{figure}

For negative $U$, the phase boundaries between the normal and inverted
states and SRA shift to lower and higher frequencies respectively, in
accordance with Eq.~(\ref{sragc}); see
Fig.~\ref{fig:omega-g-negative-u}.  This can be interpreted in terms of a
state-dependent shift of the cavity frequency, $\omega_\mp = \omega
\mp \omega_u$, as suggested by the Hamiltonian (\ref{Dickeham})
\cite{Nagy:Self,Domokos:Collective}. Within the region of overlap the
SRB phase may be stabilized, as shown in
Fig.~\ref{fig:omega-g-negative-u}. In particular, this results in a
change in both the intensity and the phase of the cavity light field
as indicated in Fig.~\ref{srasrbtran}.
In addition, in the vicinity of these transitions between SRA and SRB,
a narrow coexistence region emerges, denoted 2SRA, where both
solutions of Eq.~(\ref{szsra}) are physical; see Appendix
\ref{App:Bifurcations}.  Indeed, such coexistence phases are abundant
in the phase diagram shown in Fig.~\ref{fig:omega-g-negative-u}. For
example, as a result of the effective frequency shifts induced by
negative $U$, there is a region at low $g$ where both the normal and
inverted states coexist.  More strikingly, for $UN<-2\kappa$ in this
overlap region, there is an extension of the region of superradiant
phases to lower $g$, so that the SRB fixed point can coexist with both
the normal and the inverted states; see for example the point (d) in
Fig.~\ref{fig:omega-g-negative-u}.  In such a region, there are
multiple possible stable attractors and the ultimate fate of the
system depends on the initial conditions. In particular, this will
lead to multistability of the cavity output field.  In addition,
hysteresis can occur in these multistable regions.  For example,
increasing to a large value of $g$, and then slowly reducing the value
would lead to superradiant behavior at the point (d) shown in
Fig.~\ref{fig:omega-g-negative-u}; in contrast, slowly increasing $g$
from zero would allow the system to remain in the normal state for the
same final parameters.

It is evident from the above discussion that the  behavior of
the open system is extremely rich and is fundamentally
distinct from the equilibrium case with $\kappa=0$.  As emphasized
above and in Ref.~\cite{Keeling:Collective}, the behavior of the open
system is controlled by the stable attractors, which do not
necessarily coincide with the points of minimal free energy.  As such,
there is a crucial distinction between the $\kappa \to 0$ limit of the
dynamical system, and the equilibrium behavior at $\kappa=0$ \footnote{It is notable that in the regime where the dynamically stable
phase is SRB, the equilibrium system considered in
Ref.~\cite{Liu:Lightshift} for $\kappa=0$ is thermodynamically
unstable; the Hamiltonian is not bounded from below and the minimum
energy state occurs at infinite photon number. Such an infinite
density is unphysical, particularly in the presence of a non-vanishing
photon loss rate $\kappa$.}. In order to address experiments
with an open cavity one must consider time dependent dynamics and
$\kappa\neq 0$. In this
setting, equilibrium concepts should only be applied with caution.

\subsection{Nature of Attractors}
\label{sec:nature-attractors-u0}

\begin{figure}[!t]
  \centering
  \includegraphics[width=3.2in]{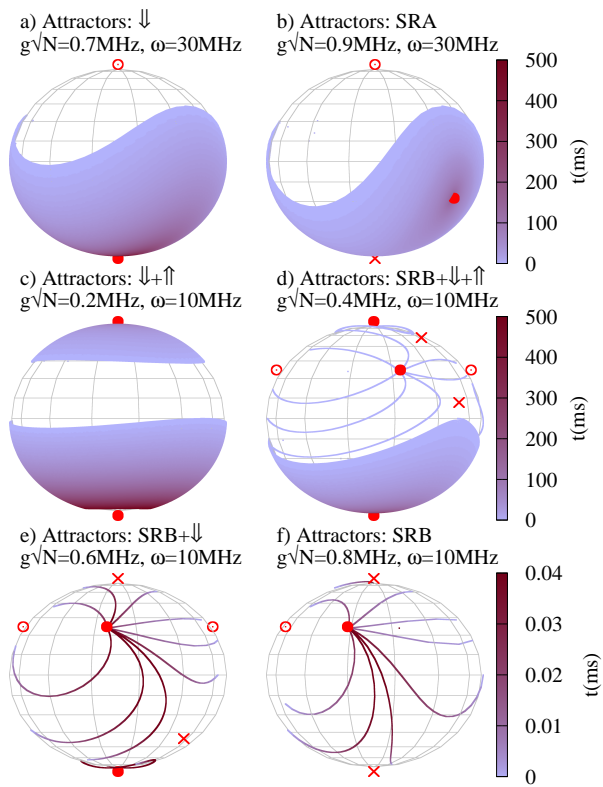}
  \caption{(color online).  
Bloch spheres with $|{\bf S}|=N/2$ 
corresponding to the points
    (a) - (f) in Fig.~\ref{fig:omega-g-negative-u} showing the fixed
    points where $\dot{\bf S}=0$ and $\dot\psi=0$. We distinguish
    between stable fixed points (filled circles), unstable fixed
    points (open circles) and hyperbolic points with one stable and
    one unstable eigenmode (crosses). The presence of more than one
    stable attractor on the Bloch sphere corresponds to a coexistence
    phase in Fig.~\ref{fig:omega-g-negative-u}.  The trajectories show
    the typical time evolution, where the initial conditions are
    chosen in order to illustrate the attractors. In particular, the
    timescale for approaching the SRA fixed point in panel (b) is
    significantly longer than the timescale for approaching the SRB
    fixed point in (f). In the language of dynamical systems the
    transition from (a) to (b) is a pitchfork bifurcation where a
    single mode goes unstable. The transition from (c) to (a) is a
    subcritical Hopf bifurcation where two modes go unstable
    simultaneously. The transition from (c) to (d)
    involves the appearance of eight additional fixed points; these
    correspond to two stable and two unstable SRB fixed points and
    four hyperbolic SRA fixed points. The transitions from (d) to (e)
    and (e) to (f) are inverse pitchfork bifurcations in which a pair
    of hyperbolic SRA fixed points coalesce at a previously stable
    fixed point.}
  \label{fig:characteristic-trajectories}
\end{figure}

In the previous section we have considered the dynamical phase diagram
with $g=g^\prime$ and $U\le 0$. Here, we provide a detailed discussion
of the nature of the long time attractors indicated in
Fig.~\ref{fig:omega-g-negative-u}. We focus on the representative
points (a) - (f) shown in Fig.~\ref{fig:omega-g-negative-u}, and chart
the associated motion of the spins on the Bloch sphere with $|{\bf
  S}|=N/2$. In Fig.~\ref{fig:characteristic-trajectories},
we show the results obtained by numerical integration of the
differential equations in Eq.~(\ref{eqmo}), using an adaptive time
step fourth order Runge-Kutta routine from the NAG
library~\cite{nag}. In addition to the characteristic trajectories, in
Fig.~\ref{fig:characteristic-trajectories} we indicate the locations
and nature of the various fixed points with $\dot\psi=0$ and $\dot{\bf
  S}=0$.  The nature of these attractors is determined by the number
of unstable eigenmodes for small fluctuations.  If no eigenmodes are
unstable, the fixed point is stable and is indicated by a filled
circle. If one eigenmode is unstable it is a hyperbolic fixed point
(or equivalently a saddle node) and is marked as a cross. If two
eigenmodes are unstable it is an unstable fixed point and is
represented by an open circle. For $\omega_0 \ll \kappa$ there are
never more than two unstable eigenmodes, meaning that the state of the
photon field rapidly comes to follow the state of the collective spin.

In order to gain some orientation we discuss the individual panels in
Fig.~\ref{fig:characteristic-trajectories}. For the parameters used in
panel (a) there is one stable attractor on the Bloch sphere
corresponding to the normal state ($\Downarrow$). This is the ultimate
fate of the system for all initial conditions as indicated in
Fig.~\ref{fig:omega-g-negative-u}; the inverted state ($\Uparrow$)
corresponds to an unstable fixed point. In passing to
Fig.~\ref{fig:characteristic-trajectories}(b), the normal state
($\Downarrow$) becomes an unstable hyperbolic fixed point and an SRA
attractor with a non-trivial magnetization $S_z$ (or its parity
symmetry partner on the other side of the Bloch sphere) governs the
long time dynamics. In contrast, in panels (c), (d) and (e) we see
multiple stable fixed points corresponding to the coexistence phases,
$\Downarrow+\Uparrow$, ${\rm SRB}+\Downarrow+\Uparrow$ and ${\rm
  SRB}+\Downarrow$ respectively; see
Fig.~\ref{fig:omega-g-negative-u}.  For these parameters, the final
state of the system depends on the initial conditions, and we only
highlight some typical trajectories. Nonetheless, the totality of
stable fixed points completely accounts for the possible asymptotic
behavior and therefore discriminates between different dynamical
phases. In panel (f) we see both stable and unstable non-trivial fixed
points corresponding to the superradiant phase SRB; see
Fig.~\ref{fig:omega-g-negative-u}. Note that in
Fig.~\ref{fig:characteristic-trajectories} we have focused on cases
with $\omega>0$. For $\omega < 0$ the fixed points can be immediately
found from the duality under the transformation given in
Eq.~(\ref{duality}). This corresponds to inverting the Bloch spheres
shown in Fig.~\ref{fig:characteristic-trajectories}.

While the attractors determine the long-time asymptotic behavior of
the system, there are cases where the evolution proceeds on
surprisingly long timescales. This may be seen in panels (a)-(d) in
Fig.~\ref{fig:characteristic-trajectories}, where the timescale for
relaxation towards the fixed point is much longer than the period of
the orbits encircling it; indeed, the latter cannot be resolved in
these panels, leading to a dense covering of the Bloch sphere. In
addition, the total time interval for approaching the SRA fixed point
in Fig.~\ref{fig:characteristic-trajectories}(b) is much longer than
that for approaching the SRB fixed point in
Fig.~\ref{fig:characteristic-trajectories}(f). We will investigate
this crucial distinction in more detail below.

\subsection{Time Evolution}
\label{sec:timeevol}         
\begin{figure}[!t]
  \centering
  \includegraphics[width=3.2in]{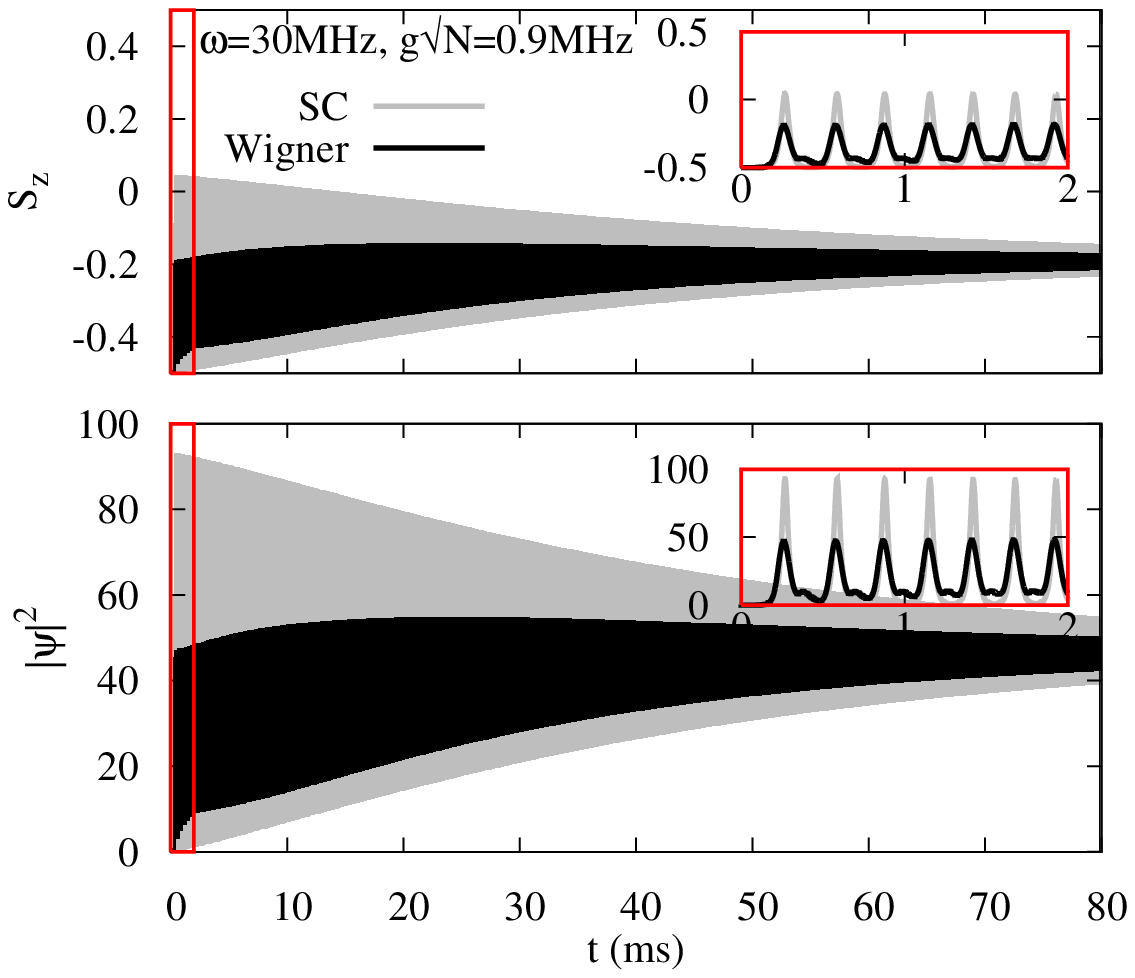}
  \includegraphics[width=3.2in]{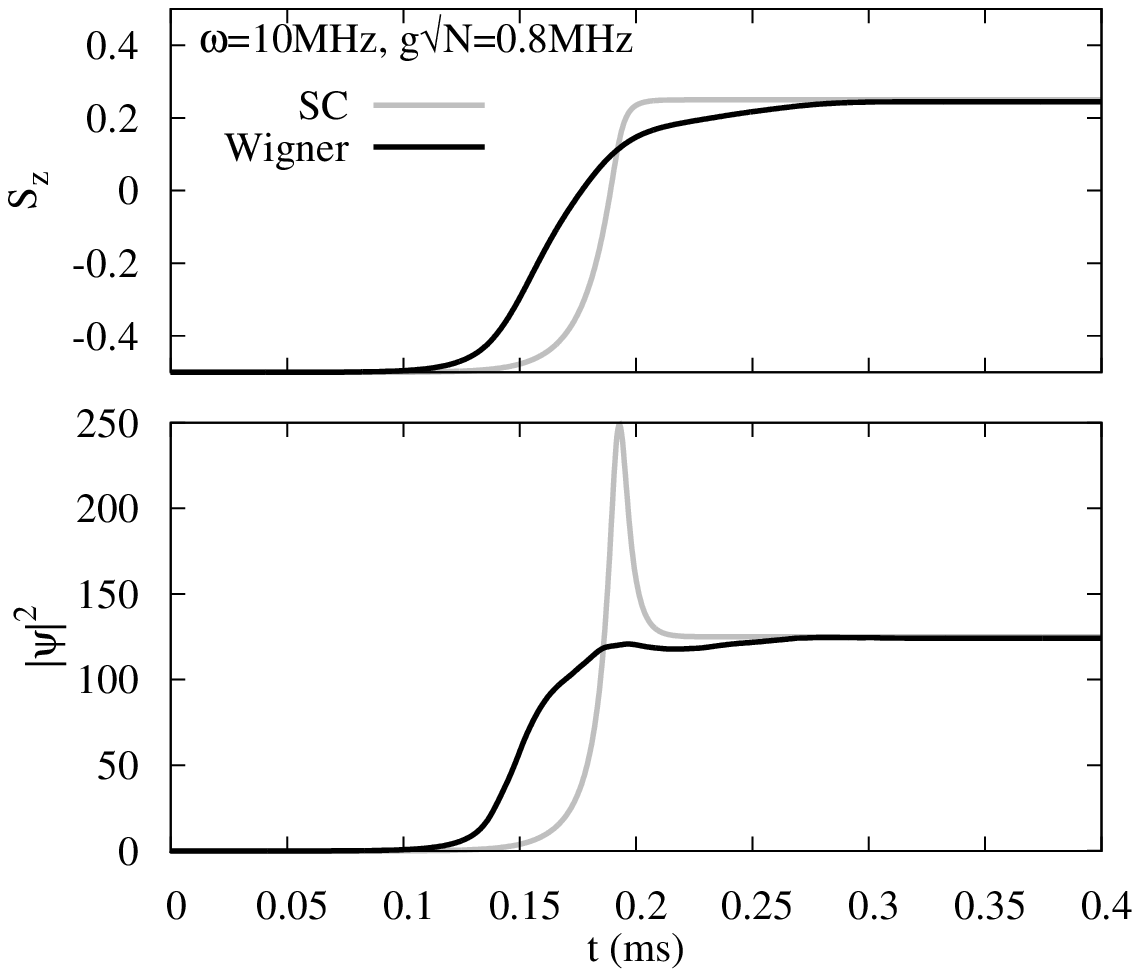}
  \caption{Time evolution in the superradiant regime with initial
    conditions that are close to the normal state ($\Downarrow$).  The
    top two panels (and insets) correspond to point (b) in
    Figs.~\ref{fig:omega-g-negative-u} and
    \ref{fig:characteristic-trajectories}.  The bottom two panels
    correspond to point (f).  In each case, the evolution of the
    semiclassical equations for a single initial condition with
    $S_x=S_y=\sqrt{N}$ is shown in gray (marked SC).  The average
    evolution for a Wigner-distributed ensemble of initial conditions
    is shown in black. The insets in the top two panels show magnified
    images of the highlighted regions.}
  \label{fig:time-evolution}
\end{figure}

In order to shed light on the dynamical distinction between the SRA
and SRB fixed points, we examine the time evolution starting near the
normal state ($\Downarrow$) for the parameters used in
Figs.~\ref{fig:characteristic-trajectories}(b) and (f) with
$\omega>0$; see Fig.~\ref{fig:time-evolution}.
By specifying an initial condition which is not a stable fixed point,
the dynamical traces correspond to a sudden quench into the
superradiant state. However, since the initial state corresponds to an
unstable fixed point, the dynamics would remain stuck in the absence
of noise or quantum fluctuations.  In practice, these inherent
fluctuations will destabilise the initial state and a non-trivial time
evolution will take place.

In order to probe the intrinsic quench dynamics we consider two
different approaches for perturbing the initial condition. The first
approach is to displace the initial state by $S_x = S_y = \sqrt{N}$,
corresponding to the characteristic size of quantum fluctuations in
the initial state; the subsequent semiclassical time evolution gives
the pale gray trajectories in Fig.~\ref{fig:time-evolution}.  The
second approach is to use a Wigner distributed ensemble of initial
conditions in order to incorporate harmonic fluctuations around the
normal state ($\Downarrow$); see for example
Refs.~\cite{itin09,Blakie:Dynamics} and Appendix \ref{App:Wigner}.  
The corresponding time dynamics is represented by the black
lines in Fig.~\ref{fig:time-evolution}. It is readily seen that both
the $\sqrt{N}$ displacement and Wigner approaches are in quantitative
agreement regarding the overall timescales for evolution towards the
SRA and SRB fixed points.  Although the amplitude of the collective
oscillations is partially washed out by the Wigner distribution of
initial spin states, oscillations of the same period nonetheless
remain in these examples.

Comparing the cases shown in Fig.~\ref{fig:time-evolution}, there is
clearly a significant difference in the relaxation timescales.  For
evolution towards the SRA fixed point shown in the upper two panels of
Fig.~\ref{fig:time-evolution}, there are oscillations at frequencies
of a few kHz, with a decay time of order $100$ms. In contrast, for
evolution towards the SRB fixed point shown in the bottom two panels
of Fig.~\ref{fig:time-evolution}, the timescale of the ``oscillation''
is similar ($\sim 0.2$ms), but only one oscillation occurs before the
steady state is reached.  As discussed in
Ref.~\cite{Keeling:Collective}, the remarkably slow dynamics near the
SRA fixed point is related to the proximity to a dynamical phase
boundary in the extended parameter space where $g$ and $g^\prime$ are
allowed to vary independently.  We will return to explore this point
further in Sec.~\ref{sec:behaviour-g-neq}.

\begin{figure}[!t]
  \centering
  \includegraphics[width=3.2in]{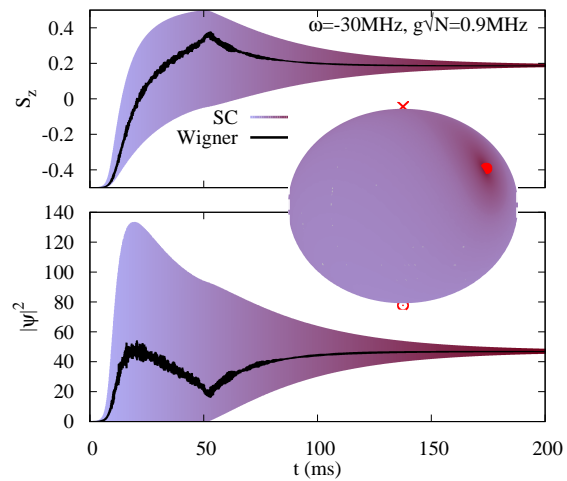}
  \caption{(color online).  
Time evolution starting close to the
    normal state ($\Downarrow$) for the parameters at point (g) in
    Fig.~\ref{fig:omega-g-negative-u}.  As in
    Fig.~\ref{fig:time-evolution}, both the semiclassical dynamics for
    a single starting point with $S_x=S_y=\sqrt{N}$ (marked SC), and a
    Wigner distributed set of initial conditions are shown.  The
    latter is indicated in black, and the semiclassical trajectory is
    shaded to match the Bloch sphere shown as an inset; for these
    initial conditions the trajectory almost covers the Bloch
    sphere. }
  \label{fig:time-evo-long-wait}
\end{figure}

Thus far we have considered the time evolution at points (b) and (f)
in Fig.~\ref{fig:omega-g-negative-u} corresponding to $\omega>0$. For
completeness we should also consider the time evolution at point (g)
with $\omega<0$. Owing to the duality in Eq.~(\ref{duality}), the
dynamics starting from $S_z=-N/2$ with $\omega>0$ is related to the
dynamics starting from $S_z=N/2$ with $\omega<0$. We should therefore
consider the quench dynamics starting from $S_z=-N/2$ and $\omega<0$
separately. This is illustrated in Fig.~\ref{fig:time-evo-long-wait},
along with the accompanying dynamics on the Bloch sphere.
It is notable that the dynamics in this regime where $\omega_-<0$ has
a remarkably long timescale. From the Bloch sphere, and the time
dependence of $S_z$, it is clear where this comes from: the trajectory
first spirals around the unstable normal state ($\Downarrow$), growing
in amplitude, until it reaches the stable manifold of the hyperbolic
inverted state ($\Uparrow$), from which it then transfers to spiral
around the stable attractor.  As such, almost the entire Bloch sphere
is covered by this trajectory, and a very long waiting time of order
$0.2$s is required before reaching the asymptotic state. In contrast
to Fig.~\ref{fig:time-evolution} the long-time asymptote is not
reached after $100$ms.  In Sec.~\ref{sec:finiteduration} we will
consider the implications of this type of behavior for the phase
diagram obtained in the experiments of Ref.~\cite{Baumann:Dicke} which
have a finite duration of order $10$ms.

\section{Growth and Decay Times and Implications for Finite 
Duration Experiments}
\label{sec:finiteduration}

In the above discussion we have seen that the timescale for reaching
the asymptotic attractors varies considerably between different points
in parameter space. In order to make contact with experiment
\cite{Baumann:Dicke} it is therefore crucial to understand how this
timescale varies throughout the phase diagram. In this section we
address this key issue.

\begin{figure}[!t]
  \centering
  \includegraphics[width=3.2in]{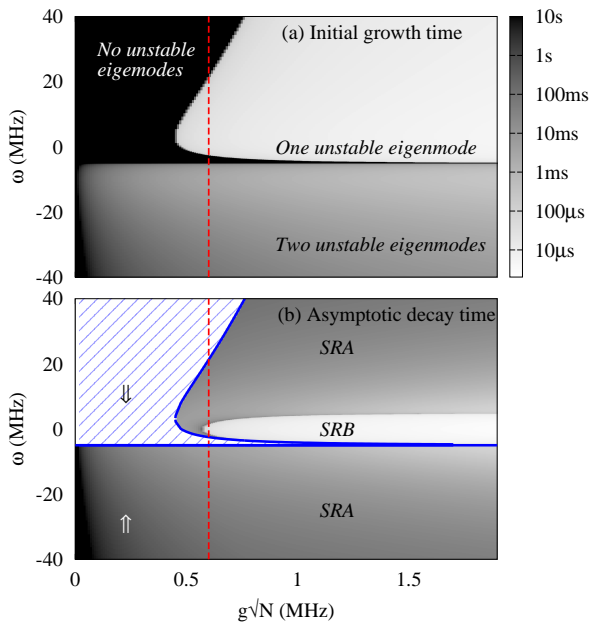}
  \caption{(color online). 
Characteristic timescales of the
    semiclassical dynamics.  We use the same parameters as in
    Fig.~\ref{fig:omega-g-negative-u} with $UN=-10$MHz corresponding
    to  the experiments of Ref.~\cite{Baumann:Dicke}. The top
    panel shows the time required for the instability of the initial
    ground state $(\Downarrow$) to develop; in regions where this state is
    stable this timescale is taken to $\infty$.  The
    lower panel shows the asymptotic timescale for approaching the final stable
    state.  In the regions of multistability, the rate of attraction
    to the final state for the given initial conditions
    ($\Downarrow$) is shown. In both panels the timescales show a strong
    variation throughout the phase diagram, with
    notable implications for finite duration experiments. A vertical
    slice along
  the red dashed line is given in Fig.~\ref{fig:growthtimes}.}
  \label{fig:instability-growth}
\end{figure}

It is evident from Fig.~\ref{fig:time-evo-long-wait} that in order to
characterize the temporal evolution we require at least two principal
timescales. The first is the timescale for departing the initial
state, and the second is the timescale for approaching the final
asymptotic attractor. Both of these timescales may be extracted by
linearizing around the initial and final states as appropriate, and
calculating the eigenvalues (frequencies) using the methods outlined
in Sec.~\ref{sec:stability} and Appendix \ref{App:linstab}. In
Fig.~\ref{fig:instability-growth}
we use these eigenvalues to plot the characteristic time for the
normal state to become unstable, if it does so, and the characteristic
decay time in the approach towards the final state; in the case of
asymptotic coexistence phases we focus on the state that is actually
reached in a quench experiment that starts close to the normal state
($\Downarrow$). It is clear from Fig.~\ref{fig:instability-growth}
that both of these fundamental timescales vary significantly
throughout the phase diagram.  In particular, in the region where
$\omega_-<0$, both the timescale for the initial destabilization of
the normal state and the timescale for decay towards the asymptotic
state are increased.  The combination of these two timescales provides
a lower bound on the overall duration of the intrinsic dynamics. For
$\omega_-<0$ we therefore expect a slower approach to the asymptotic
regime. In order to gain a better handle on this issue, we provide an
analytic discussion of the constituent timescales below. We begin in
Sec.~\ref{gr} with an analysis of the initial growth times before
examining the final asymptotic decay times in Sec.~\ref{dr}.  In
Sec.~\ref{intermediatemap} we then consider the implications for
experiments which monitor the photon intensity over a finite time
interval.

\subsection{Growth and Decay Times}
\subsubsection{Growth Times}
\label{gr}
The initial growth time in Fig.~\ref{fig:instability-growth}(a)
can be understood and estimated analytically,
by using the linearization discussed in Sec.~\ref{sec:stability}.
Considering Eq.~(\ref{equalgquart}) for the normal state with $S_z=-N/2$
the eigenvalues $\eta$ obey:
\begin{equation}
  [(\eta + i\kappa)^2 - \omega_-^2](\eta^2 - \omega_0^2) 
  - 4 \omega_0 \omega_- g^2 N  = 0,
 \label{eq:2}
\end{equation}
where $\omega_-\equiv \omega-UN/2$. To
obtain the growth rate for a given set of parameters we must find the
solution of Eq.~(\ref{eq:2}) with the largest positive imaginary part,
$\eta^{\prime\prime}$. Solving Eq.~(\ref{eq:2}) numerically we plot
the corresponding growth time $1/\eta^{\prime\prime}$ in
Fig.~\ref{fig:instability-growth}(a). It is evident that there are
distinct growth times in the top and bottom portions of
Fig.~\ref{fig:instability-growth}(a) separated by the critical line
$\omega_-=0$. The origin of this distinction may be traced to a change
in behavior of the root structure of Eq.~(\ref{eq:2}). For the
parameters shown in Fig.~\ref{fig:instability-growth}(a), when
$\omega_->0$ a single one of the four roots goes unstable, whilst for
$\omega_-<0$ two roots with same imaginary parts go unstable
simultaneously. In the language of dynamical systems, the first
scenario corresponds to a pitchfork bifurcation, whilst the latter
corresponds to a Hopf bifurcation \cite{strogatz94}.
\begin{figure}[!t]
  \centering
  \includegraphics[width=3.2in]{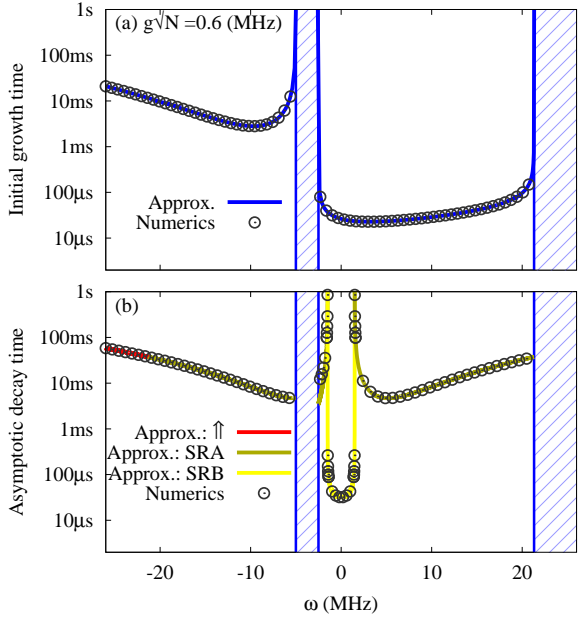}
  \caption{(color online). 
Vertical slice through
    Fig.~\ref{fig:instability-growth} with $g\sqrt{N}=0.6$MHz.  (a)
    Initial growth time obtained by linear stability analysis around
    the normal ($\Downarrow$) state. The open circles show the exact
    semiclassical results obtained numerically from the quartic
    Eq.~(\ref{eq:2}). The solid line corresponds to the approximation
    given in Eq.~(\ref{nusub}) where the plus sign corresponds to the
    unstable mode.  (b) Asymptotic decay time obtained by linear
    stability analysis around the appropriate final asymptotic
    state. The open circles correspond to the eigenvalue of
    $|\eta\tilde{\mathbb I}-\tilde{\mathbb M}|=0$ with the largest
    imaginary part, where $\tilde{\mathbb M}$ is given by
    Eq.~(\ref{m4}). The solid gold (mid gray) line corresponds to the
    imaginary part of Eq.~(\ref{eq:7}) and the solid yellow (light
    gray) line corresponds to the exact result in Eq.~(\ref{eq:9}). In
    both panels, the shaded regions indicate where the normal state
    ($\Downarrow$) is stable. Both timescales show a strong dependence
    on the parameters. }
  \label{fig:growthtimes}
\end{figure}

In order to gain a quantitative handle on the observed growth times we
may exploit the small parameter $\omega_0/\kappa$ in order to find
analytic approximations for $\eta$.  Anticipating that $|\eta|\sim
\omega_0\ll \kappa$ we neglect $\eta$ in the combination $\eta+i\kappa$.
This yields a quadratic equation which provides
\begin{equation}
  \label{eq:3}
  \eta \approx \pm \sqrt{\omega_0^2 - 
    \frac{4 \omega_0 \omega_-g^2N}{\omega_-^2+\kappa^2}}
  = \pm \omega_0 \sqrt{1 - \left(\frac{g}{g_a^{-}}\right)^2},
\end{equation}
where $g_a^{-}$ is given in Eq.~(\ref{sragc}) and the plus sign
corresponds to the growing mode. For $g > g_a^{-}$, $\eta$ is
imaginary, and the characteristic growth rate is indeed of order
$\omega_0$ as we assumed. However, for $\omega_-<0$ it is evident from
the first form of Eq.~(\ref{eq:3}) that $\eta \in \mathbb{R}$ and
therefore Eq.~(\ref{eq:3}) cannot apply in the lower region of
Fig.~\ref{fig:instability-growth}(a).  To fully describe the observed
behavior it is necessary to carry out the expansion of $\eta$ to
higher order in $\omega_0/\kappa$. Parameterizing $\eta=\pm
\omega_0\sqrt{1-(g/g_a^-)^2}+\delta_\pm$, where $\delta_\pm\ll
\omega_0\ll \kappa$, and substituting into Eq.~(\ref{eq:2}) one
obtains
\begin{equation}
  \eta\approx \pm \omega_0 \sqrt{1 - \left(\frac{g}{g_a^{-}}\right)^2}-\frac{4i\kappa\omega_0\omega_-g^2N}{(\omega_-^2+\kappa^2)^2}.
\label{nusub}
\end{equation}

As shown in Fig.~\ref{fig:growthtimes}(a), the root with the plus sign
in Eq.~(\ref{nusub}) accurately reproduces the exact growth times
observed in Fig.~\ref{fig:instability-growth}(a). In particular, for
$\omega_->0$ one may neglect the last term in Eq.~(\ref{nusub}), but
for $\omega_-<0$ it is essential to retain this contribution as the
leading term no longer corresponds to a decay rate. In the former case
where $\omega_->0$, the decay rate is typically of order
$\omega_0=0.05$MHz; this corresponds to a timescale of 20$\mu$s, in
agreement with the upper region of
Fig.~\ref{fig:instability-growth}(a) and the central portion of
Fig.~\ref{fig:growthtimes}(a).  In the latter case, where
$\omega_-<0$, the decay rate is of order $\omega_0^2/\kappa\simeq
0.3$kHz, where we use the functional dependence of the critical
coupling in Eq.~(\ref{sragc}). This corresponds to a much longer
timescale of order $3$ms, in agreement with the lower region of
Fig.~\ref{fig:instability-growth}(a) and the left hand side of
Fig.~\ref{fig:growthtimes}(a). In fact the exact timescales diverge on
approaching the phase boundaries where the $\Downarrow$ state is
stable; see Fig.~\ref{fig:growthtimes}(a). This may also be
interpreted as critical slowing down \cite{Keeling:Collective}. For
example, approaching the left region of $\Downarrow$ in
Fig.~\ref{fig:growthtimes}(a) from the right, one has
$\eta\sim\sqrt{g_a^--g}$; this is analogous to an equilibrium mean
field exponent $z\nu=1/2$. On the other hand, approaching the left
region of $\Downarrow$ from the left, one has $\eta\sim
(\omega-UN/2)$.  This latter behavior arises from the second term in
Eq.~(\ref{nusub}) and only exists in the open system with $\kappa\neq
0$; it is an analogous to a critical exponent $z\nu=1$. A divergent
growth time may also be seen in the lower left corner of
Fig.~\ref{fig:instability-growth}(a), since $g\rightarrow 0$ in the
second term of Eq.~(\ref{nusub}). For recent discussions of critical
behavior in driven open cavities see
Refs.~\cite{Szirmai:Noise,Oztop:Driven,Nagy:Exponent,Bastidas:Noneq}.

\subsubsection{Decay Times}
\label{dr}
Turning to the asymptotic decay time in
Fig.~\ref{fig:instability-growth}(b), we consider the approach towards
the three stable fixed points, $\Uparrow$, SRA and SRB which differ
from the initial state, $\Downarrow$. In order to extract the
associated timescales we must linearize around the asymptotic fixed
points and find the eigenvalue with the smallest imaginary part.

For decay towards the inverted state ($\Uparrow$), we may invoke our
previous result in Eq.~(\ref{nusub}) with the replacement $\omega_- \to
\omega_+$. In order to extract the timescale governing the
approach towards SRA we must linearize around this fixed
point. Following the general approach used in Sec.~\ref{sec:stability}
this yields the characteristic equation
\begin{equation}
  \label{eq:6}
  [(\eta+i\kappa)^2 - \tilde\omega^2]\left[ \eta^2 -
    \left(\frac{{\tilde\omega}_0 N}{2 S_z}\right)^2\right]
  = \\
  - \frac{2 \tilde{\omega}_0 \tilde{\omega}}{S_z} |2 g S_z - U \psi S_x
 |^2
\end{equation}
where $\tilde\omega\equiv \omega +US_z$ and $\tilde{\omega}_0 \equiv
\omega_0 + U |\psi|^2$ are useful variables suggested by
Eq.~(\ref{eqmo}); for more details see Appendix \ref{App:linstab} and
Eq.~(\ref{appsrapoly}).  The ultimate timescale controlling the decay
towards the fixed point is governed by the slowest roots of
Eq.~(\ref{eq:6}).  Anticipating that these have $|\eta|\sim
\tilde\omega_0N/2S_z\ll \kappa$ we may once again neglect the term
$\eta$ in the combination $\eta+i\kappa$. This yields a quadratic
equation for $\eta$ with solutions $\eta=\pm \eta_0$ where $\eta_0\in
{\mathbb R}$.  In order to refine this approximation we parameterize
$\eta=\pm \eta_0+\delta_\pm$ where $\delta_\pm\ll \omega_0\ll \kappa$,
and substitute into Eq.~(\ref{eq:6}). Retaining terms up to linear
order in $\delta$ one obtains
\begin{equation}
\label{eq:7}
\eta\approx \pm \eta_0+\frac{2i\kappa\tilde{\omega}_0\tilde{\omega} |2g S_z - U S_x \psi|^2}{(\tilde{\omega}^2+\kappa^2)^2 S_z}.
\end{equation}
In the limit $U=0$ one finds ${\rm Im}(\eta) \approx - \kappa
\omega_0^2 / (\omega^2+\kappa^2)$, where we use Eq.~(\ref{sraquad}) to
substitute for $S_z$.  This agrees with our previous findings
\cite{Keeling:Collective}, and yields a characteristic decay rate of
order $\omega_0^2 / \kappa \simeq 0.3$kHz. This is consistent with the
$3$ms timescale found in Figs.~\ref{fig:instability-growth}(b) and
\ref{fig:growthtimes}(b). In addition, in the limiting case of the SRA
phase where $\psi=0$ and $S_z=-N/2$, we recover the characteristic
frequencies of the normal ($\Downarrow$) state as given by
Eq.~(\ref{nusub}). As shown in Fig.~\ref{fig:growthtimes}(b),
Eq.~(\ref{eq:7}) accurately reproduces the results obtained by direct
numerical solution of Eq.~(\ref{eq:6}).

For decay towards SRB, the results are much simpler. In this case the
characteristic frequencies satisfy
\begin{equation}
\left[\eta(\eta+i\kappa) - 4 gU\psi_2S_y \right]^2=0,
\label{srbchar}
\end{equation}
where $\psi\equiv \psi_1+i\psi_2$; see Appendix \ref{App:linstab}. In
the case of the stable SRB fixed points discussed in
Sec.~\ref{sec:srb}, where
$(\psi,S_y)=\pm(i\sqrt{(-\omega_0/U)},|S_y|)$, one readily obtains
exact results for the repeated roots of Eq.~(\ref{srbchar}):
\begin{equation}
  \label{eq:9}
  \eta = - i \frac{\kappa}{2} \pm \sqrt{- \frac{\kappa^2}{4} 
    + 2\omega_0 \kappa \sqrt{ \frac{g^2-g_b^2}{g_b^2} } },
\end{equation}
where we use the fact that
$S_y^2=\omega_0\kappa^2(g^{-2}-g_b^{-2})/4U$. The decay towards SRB is
governed by the slowest mode corresponding to the positive root in
Eq.~(\ref{eq:9}).  Since $\omega_0\ll \kappa$ we may Taylor expand
this root to obtain ${\rm Im}(\eta)\simeq
-2\omega_0\sqrt{g^2/g_b^2-1}+{\mathcal O}(\omega_0^2/\kappa)$.  As
shown in Fig.~\ref{fig:growthtimes}(b), the exact analytic result
(\ref{eq:9}) is in agreement with the numerical solution of
Eq.~(\ref{srbchar}) as required.

From the above analysis we see that the characterstic decay rates
towards SRA and SRB are of order $\omega_0^2/\kappa$ and $\omega_0$
respectively. These correspond to decay times of the order of $3$ms
and $20\mu$s which yields a faster approach towards SRB in comparison
to SRA. This is confirmed by the typical trajectories on the Bloch
sphere as shown in Figs.~\ref{fig:characteristic-trajectories}(b) and
(f). It is readily seen that many more orbits are executed in reaching
the SRA fixed point.  We see that the SRA and SRB attractors differ in
their dynamic characteristics, in addition to their steady state
forms.

\subsection{Photon Intensity Map Extracted at Intermediate Times and
  Implications for Finite Duration Experiments}
\label{intermediatemap}

\begin{figure}[!t]
  \centering
  \includegraphics[width=3.2in]{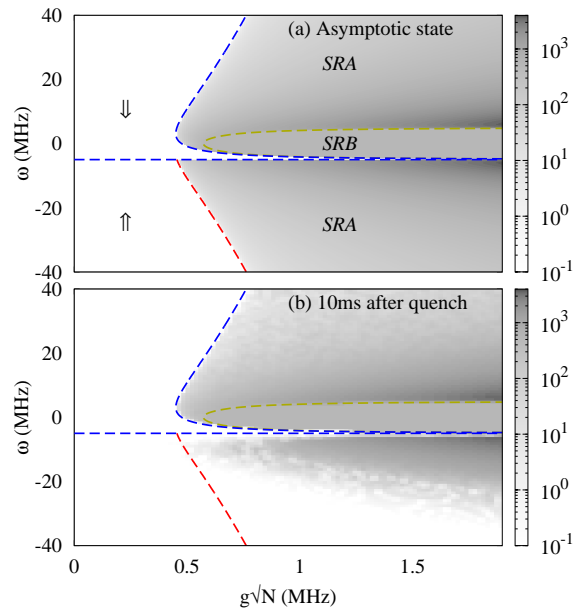}
  \caption{(color online).  
Photon intensity maps showing $|\psi|^2$
    after two different time intervals, with initial conditions that
    are close to the normal state ($\Downarrow$) with
    $S_x=S_y=\sqrt{N}$.  We use the same parameters as in the second
    panel of Fig.~\ref{fig:omega-g-negative-u} with $UN=-10$MHz,
    corresponding to the experiments of Ref.~\cite{Baumann:Dicke}. (a)
    Intensity map obtained in the final asymptotic state with
    $t\rightarrow\infty$, showing the distinct regions of SRA, SRB and
    $\Uparrow$.  (b) Intensity map obtained after $10$ms showing good
    qualitative agreement with the asymptotic attractors in panel (a),
    but with a slower approach in the SRA regions. A cross section of
    panel (a) with $g\sqrt{N}=1.0$MHz is provided in
    Fig.~\ref{srasrbtran}.}
  \label{fig:quench-map}
\end{figure}

Having examined the characteristic growth and decay times across the
phase diagram, we now consider the consequences for finite duration
experiments. In particular, for $\omega_-<0$, the semiclassical
dynamics predicts relatively long growth times and comparably long
approach times, as shown in Fig.~\ref{fig:instability-growth}.  In
Fig.~\ref{fig:quench-map} we compare the resulting photon intensity
map obtained after a hypothetical infinite duration experiment, to
that obtained after $10$ms. It is readily seen that the lower region,
corresponding to $\omega_-<0$, has not fully reached the asymptotic
regime.

\begin{figure}[!t]
  \centering
\includegraphics[width=3.2in]{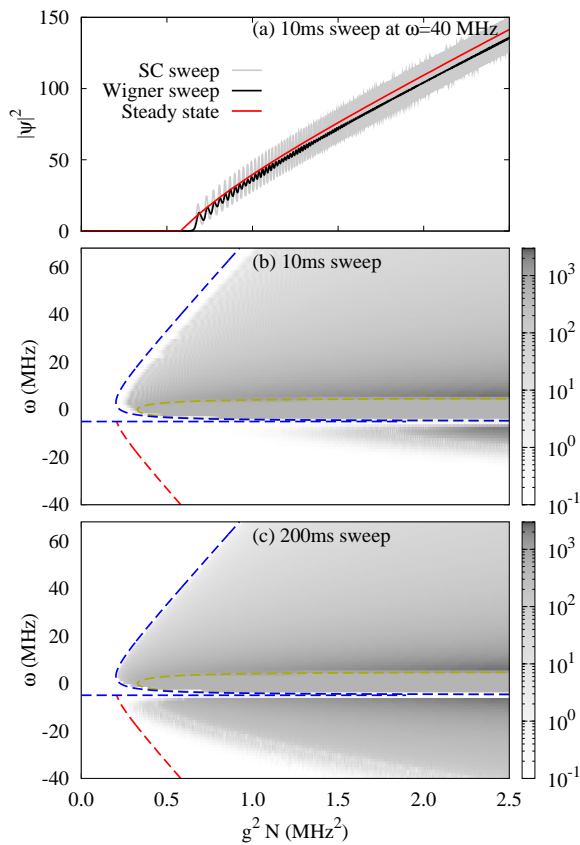}
\caption{(color online).  
$|\psi|^2$ found by increasing $g^2 \propto
  t$, and recording the value achieved as a function of time.  We use
  the same parameters as in Fig.~\ref{fig:omega-g-negative-u} with
  $UN=-10$MHz, corresponding to the experiments of
  Ref.~\cite{Baumann:Dicke}. The sweep is chosen so that $g^2 N =
  (t/t_0)\times 2.5\text{MHz}^2$, where $t_0=10$ms in panels (a) and
  (b) and $t_0=200$ms in panel (c). The top two panels correspond to
  the experimental sweep duration used in Ref.~\cite{Baumann:Dicke}.
  (a) Comparison of the steady state value of $|\psi|^2$ with that
  obtained by semiclassical evolution (marked SC) and Wigner
  distributed initial conditions, for the value of $g\sqrt{N}$ reached
  at a given time with $\omega=40$MHz.  (b) Photon intensity map
  obtained after a $10$ms sweep. (c) Photon intensity map obtained
  after a $200$ms sweep.  In comparing these figures to Fig. 5 of
  Ref.~\cite{Baumann:Dicke} one should note that the vertical scale on
  our intensity plots is the photon frequency $\omega$ appearing in
  Eq.~(\ref{Dickeham}). In comparison, Ref.~\cite{Baumann:Dicke} use
  the detuning of the pump from the bare-cavity frequency as the
  vertical scale, hence the inverted and shifted axis.}
  \label{fig:sweep-dynamics}
\end{figure}

In order to make close contact with the experiment of
Ref.~\cite{Baumann:Dicke}, we should also incorporate the details of
their data aquisition scheme. In particular, the photon intensity map
is obtained by increasing the laser intensity over a $10$ms interval
and recording the photon intensity during this period.  This procedure
is then repeated for other detunings and an intensity map is
generated.  In order to facilitate a direct comparison, we incorporate
the effects of the sweep in our numerical simulations, where we take
$g^2 \propto t$. In Fig.~\ref{fig:sweep-dynamics}(a) we show how for
one value of $\omega$, $|\psi|^2$ evolves with increasing
matter--light coupling where we take $g^2N = (t/t_0) \times 2.5 {\rm
  MHz}^2$, where $t_0=10$ms in Figs.~\ref{fig:sweep-dynamics}(a) and
\ref{fig:sweep-dynamics}(b), and $t_0=200$ms in
Fig.~\ref{fig:sweep-dynamics}(c).  As discussed in
Sec.~\ref{sec:timeevol} we present both a single semiclassical
trajectory with $S_x=S_y=\sqrt{N}$, and the results of Wigner
distributed initial conditions. As found previously, it is readily
seen from Fig.~\ref{fig:sweep-dynamics}(a) that quantum fluctuations
reduce the oscillations in the photon intensity, but that the overall
dependence conforms to the semiclassical analysis. Moreover, for this
set of parameters, the results of the $10$ms sweep are also in good
agreement with the steady state photon intensity. We may therefore use
the semiclassical approach to map out the resultant phase diagram.

A notable finding is that in other regions of the phase diagram,
sweeps longer than $10$ms may be required in order to reveal
signatures of the asymptotic attractors. To see this more clearly, in
Figs.~\ref{fig:sweep-dynamics}(b) and (c) we show the results of the
semiclassical evolution in Eq.~(\ref{eqmo}), with $\sqrt{N}$
displacement of the initial state, and a sweep profile.
It is readily seen that for $\omega_->0$, the results of the $10$ms
sweep are already quite close to the long time asymptotic state.
However, as can be anticipated by the large instability growth and
asymptotic approach times for $\omega_-<0$, the normal state
($\Downarrow$) persists. Despite its ultimate instability, the $10$ms
is insufficient for the instability to grow. In contrast, with a sweep
duration of $200$ms, one can see the instability of the normal state
in this region, although the final asymptotic state of the
semiclassical dynamics has not been reached.
\section{General Phase Diagram}
\label{sec:genpd}
Having discussed the phase diagram and the collective dynamics for the
experimentally explored case with $g=g^\prime$ and $U<0$
\cite{Baumann:Dicke}, we now consider the broader parameter space. In
Sec.~\ref{sec:phase-diagram-u0} we consider the case with $g=g^\prime$
and $U>0$, and in Sec.~\ref{sec:behaviour-g-neq} we examine the case
with $g\neq g^\prime$. A notable feature of both of these cases are
parameter regimes in which no stable fixed point exists, and for which
persistent oscillations arise. 

\subsection{Phase diagram for $g=g^\prime$ and $U>0$}
\label{sec:phase-diagram-u0}

\begin{figure}[!t]
  \centering
  \includegraphics[width=3.2in]{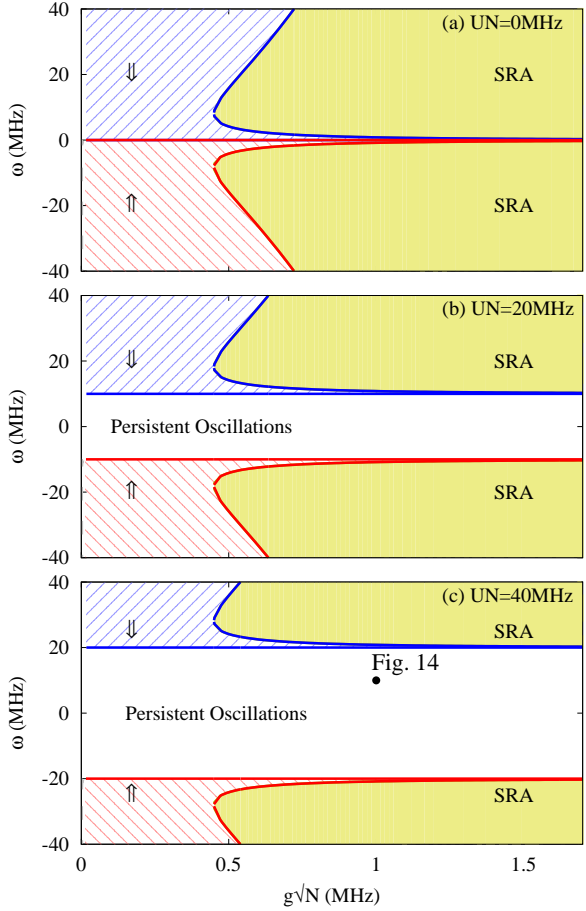}
  \caption{(color online). 
Dynamical phase diagram as a function of
    $\omega$ and $g=g^\prime$ for $U\ge 0$. The phase boundaries at
    $\omega_\pm=0$ separate with increasing $U$ and a regime of
    persistent oscillations emerges. The dynamics in this regime is
    shown in Fig.~\ref{fig:equal-g-limit-cycle}. In contrast to
    Fig.~\ref{fig:omega-g-negative-u}, the SRB phase is absent here 
since $U\ge 0$.}
  \label{fig:omega-g-positive-u}
\end{figure}

The sign of $U$ can be varied by switching between red and blue
detuning of the cavity light field with respect to the atomic
transition. This may be seen from the derivation of the effective
Dicke model Hamiltonian, as outlined in Sec.~\ref{Sect:Exp} and
Appendix~\ref{sec:depart-from-gener}.  In
Fig.~\ref{fig:omega-g-positive-u} we show the resulting phase diagram
with $g=g^\prime$ and $U>0$; this is the analog of the phase diagram
shown in Fig.~\ref{fig:omega-g-negative-u}.

\begin{figure}[!t]
  \centering
  \includegraphics[width=3.2in]{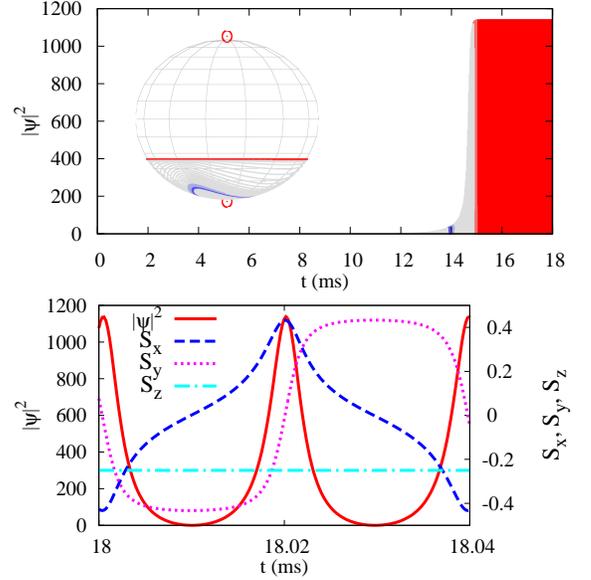}
  \caption{(color online).  
Persistent oscillations at $\omega=10$MHz,
    $UN=+40$MHz, $g\sqrt{N}=1$MHz starting close to the normal state
    ($\Downarrow$) with $S_x=S_y=\sqrt{N}$.  The top panel shows the
    attraction towards persistent oscillations, illustrating the
    transient behavior at short times, and the persistent oscillations
    at later times.  The inset shows the same data on the Bloch sphere
    using the same shading scheme. For all times after $\sim15$ms, the
    trajectory on the Bloch sphere is restricted to a circle at
    constant polar angle.  The lower panel shows the time dependence
    of $|\psi|^2$ and $\vect{S}$ in the persistent oscillation
    regime.}
  \label{fig:equal-g-limit-cycle}
\end{figure}

With $U>0$, the phase boundaries at $\omega_\pm=0$ shift in the
opposite direction to those obtained with $U<0$.  As such, the
boundaries separate, rather than overlap, as shown in
Fig.~\ref{fig:omega-g-positive-u}. Instead of finding coexistence
phases, as found in Fig.~\ref{fig:omega-g-negative-u}, a regime of
persistent oscillations emerges as shown in white in
Fig.~\ref{fig:omega-g-positive-u}.  In this region, no steady state is
ever reached, and the photon number continues to oscillate
periodically at long times. This is illustrated in
Fig.~\ref{fig:equal-g-limit-cycle}.

For these persistent oscillations with $g=g^\prime$ it is possible to
characterize their behavior analytically.  In this case the emergent
state is a limit cycle \cite{Keeling:Collective}. To see this one may
note from Fig.~\ref{fig:equal-g-limit-cycle} that the asymptotic
behavior has constant $S_z$, and in fact $S_z=-\omega/U$.  From the
equations of motion in Eq.~(\ref{eqmo}) we see that $\dot S_z = -ig
(\psi + \psi^\ast)(S^+ - S^-)$. However it also clear from
Fig.~\ref{fig:equal-g-limit-cycle} that $S_y \neq 0$ and so $S^+ \neq
S^-$.  We therefore require ${\rm Re}(\psi) =0$, as found in the SRB
steady state.  With these conditions on $S_z$ and $\psi$,
Eq.~(\ref{eqmo}) simplifies. Writing $S^- = r e^{-i \theta}$, where $r^2= N^2/4 -
\omega^2/U^2$ is a constant of the motion, one obtains
\begin{equation}
  \label{eq:13}
  \dot{\theta} = \omega_0 + U |\psi|^2, \quad
  \dot{\psi} + \kappa \psi = -2 i g r \cos \theta.
\end{equation}
This pair of coupled first order equations describes the exact
dynamics of the persistent oscillations.  Since the motion is in a
two-dimensional plane, the attractor is a simple limit cycle
\cite{strogatz94}.  In Eq.~(\ref{eq:13}), the phase angle $\theta$
continually increases, but has alternate fast and slow regions; the
motion is faster when $|\psi|$ is larger as may be seen in
Fig.~\ref{fig:equal-g-limit-cycle}.  Such behavior is analogous to a
damped driven pendulum. In fact, for $\kappa \gg \omega_0 + U
|\psi|^2$, one may adiabatically eliminate $|\psi|$ to obtain
\begin{math}
  \dot\theta = (\omega_0 + \lambda)
  + 
  \lambda \cos(2 \theta)
\end{math}
where $ \lambda = U g^2 r^2/2 \kappa$. This is the equation of motion for 
a damped driven pendulum, and since $\omega_0 > 0$ it is driven
above the threshold required for persistent oscillations.

\subsection{Phase Diagram for $g \neq g^\prime$}
\label{sec:behaviour-g-neq}

Up until now, we have mainly restricted our discussion to the
experimentally realized case where one necessarily has $g=g^\prime$
\cite{Baumann:Dicke}.  However, there are important reasons to explore
what happens when this condition is relaxed. In particular, there are
a number of phase boundaries in the extended $g, g^\prime$ parameter
space, and these can be rather close to the experimental situation
with $g=g^\prime$.  As highlighted in Ref.~\cite{Keeling:Collective},
proximity to these phase boundaries is instrumental in explaining
regions of slow decay in the $g=g^\prime$ dynamics.  Secondly, the
proposal of Dimer {\em et al} \cite{Dimer:Proposed} considers a Raman
scheme, rather than Rayleigh scattering, and involves different
hyperfine atomic states.  In this setting, separate tuning of $g$ and
$g^\prime$ could be acheived by using circularly polarized pump beams
in a ring cavity \cite{Dimer:Proposed}. For these reasons, we consider
the behavior for $g\neq g^\prime$.

\begin{figure}[!t]
  \centering
  \includegraphics[width=3.2in]{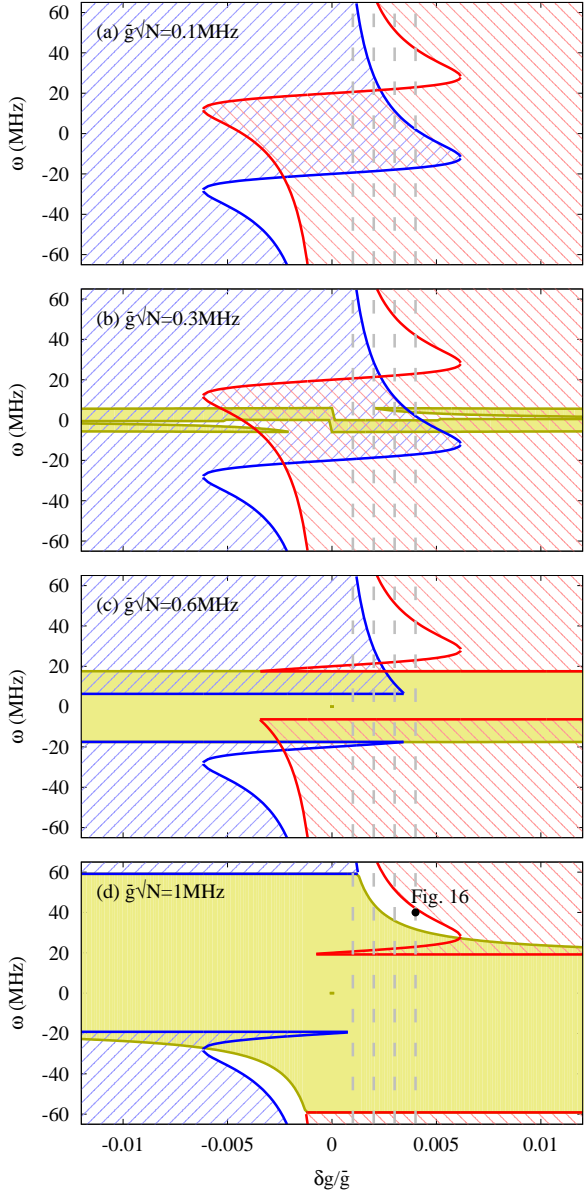}
  \caption{(color online). 
Dynamical phase diagram at $UN=-40$MHz, as
    a function of $\delta g \equiv g^\prime-g$ and $\omega$ for a
    number of values of $\bar{g} = \frac{1}{2} (g+g^\prime)$. Vertical
    dashed lines indicate cuts shown in
    Fig.~\ref{fig:omega-g-negative-u-varying-ratio}.  }
  \label{fig:omega-g-negative-u-vs-dg}
\end{figure}

\begin{figure}[!t]
  \centering
  \includegraphics[width=3.2in]{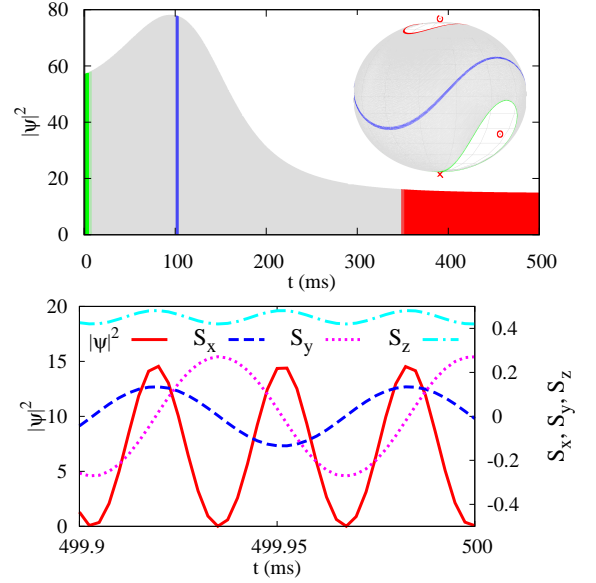}
  \caption{(color online). 
Persistent oscillations for $UN=-40$MHz,
    $\omega=40$MHz, $g \sqrt{N}=0.998$MHz and $g^\prime
    \sqrt{N}=1.002$MHz, corresponding to $\delta g/\bar{g} = 0.004$ or
    $g^\prime/g=1.004$.  The panels mirror those in
    Fig.~\ref{fig:equal-g-limit-cycle}.  The shading in the top panel
    match those in the inset, highlighting the initial and
    intermediate trajectories, and the final persistent oscillations.}
  \label{fig:non-equal-g-limit-cycle}
\end{figure}

In Fig.~\ref{fig:omega-g-negative-u-vs-dg}
we set $UN=-40$MHz, and explore deformations by $\delta g \equiv g^\prime
- g$ at four different values of fixed $\bar{g} \equiv \frac{1}{2}
(g+g^\prime)$.
There are three key aspects to note. The first concerns the existence
of non-trivial phase boundaries in proximity to the $\delta g=0$, or
$g=g^\prime$ axis. In particular, as one transits along the $\delta
g=0$ axis in Fig.~\ref{fig:omega-g-negative-u-vs-dg}(d) there are two
distinct scenarios depending on whether $|\omega|>|\omega_u|$ or
$|\omega|<|\omega_u|$, where $\omega_u\equiv UN/2$.  In the former
case there is a proximate phase boundary for small $\delta g/{\bar
  g}$, and associated critical slowing down.  In the latter case the
closest phase boundary is horizontal and is therefore not crossed by
changing $\delta g$. Therefore, for a broad range of
$|\omega|<|\omega_u|$ one may avoid close proximity to a phase
boundary and the associated critical slowing down.  This
$\omega$-dependence of the emergent timescales is confirmed in
Fig.~\ref{fig:time-evolution}. The second notable feature in
Fig.~\ref{fig:omega-g-negative-u-vs-dg}(d) is that the SRA and SRB
phases, which are distinct for $g=g^\prime$, are continuously
connected for $g\neq g^\prime$. This may be traced to the lack of
factorization of the equation of motion for $\dot S_z$, which
simplifies at $g=g^\prime$ to $\dot
S_z=-ig(\psi+\psi^\ast)(S^+-S^-)$. The third notable feature is that
there are again regions of persistent oscillations, shown in white,
where no stable fixed points exist. As shown in
Fig.~\ref{fig:non-equal-g-limit-cycle},
the detailed dynamics differs from the persistent oscillations
discussed in Sec.~\ref{sec:phase-diagram-u0}, as may be seen from the
feature that $S_z$ is no longer constant.

In general it may be difficult to gain a purely analytic handle on the
equations of motion in Eq.~(\ref{eqmo}) when $g\neq g^\prime$ and $U$
is present. However, in the adiabatic limit with
$\kappa\rightarrow\infty$ one may eliminate the photons and consider
the dynamics of the spins alone. In this limit the equations of motion reduce 
to the following form
\begin{equation}
  \label{eq:14}
  \dot{\vect{S}} = \left\{ \vect{S}, H \right\}
  -{\bf D}_1({\bf S})
  -{\bf D}_2({\bf S}),
\end{equation}
where the effective Hamiltonian is given by 
\begin{equation}
  H=\omega_0 S_z - \frac{\tilde{\omega} (G_+ S_x^2 + G_- S_y^2)}{\kappa^2 + \tilde{\omega}^2},
\label{hadia}
\end{equation}
with $\tilde{\omega} \equiv \omega + U S_z$. The additional contributions
\begin{equation}
\begin{aligned}
{\bf D}_1 & =\frac{2 \kappa \Gamma \vect{S}\times(\vect{S}\times \hat{\vect{z}})}{\kappa^2 + \tilde{\omega}^2}, \\
{\bf D}_2 & =\frac{2 \kappa^2 U (G_+ S_x^2 + G_- S_y^2)}{(\kappa^2 + \tilde{\omega}^2)^2} 
  \vect{S}\times \hat{\vect{z}},
\end{aligned}
\end{equation}
are damping terms with $G_\pm \equiv (g^\prime \pm g)^2$ and $\Gamma
\equiv g^{\prime 2} - g^2$. The existence of the photon leakage
$\kappa$ therefore means that there are two non-Hamiltonian terms in
the effective spin dynamics. In the special case where $U=0$ the term
${\bf D}_1$ has the same form as the damping in the
Landau-Lifshitz-Gilbert equations \cite{Landau:LLG,Gilbert:LLG} and
${\bf D}_2={\bf 0}$. The former tries to drive the system toward
either the normal or inverted states. In this $U=0$ limit the
Hamiltonian contribution in Eq.~(\ref{hadia}) also reduces to an
effective Lipkin--Meshkov--Glick (LMG) Hamiltonian
\cite{Lipkin:LMG1,Lipkin:LMG2,Lipkin:LMG3,Morrison:Coll}.

A notable aspect of the adiabatic limit with
$\kappa\rightarrow\infty$, is that the resulting dynamics in
Eq.~(\ref{eq:14}) resides solely within the two-dimensional surface of
the Bloch sphere. The Poincar\'e-Bendixson theorem \cite{strogatz94}
therefore excludes the possibility of chaotic attractors.  In
contrast, for $\kappa=0$, the conservative Dicke model with
$g=g^\prime$ and $U=0$ is known to exhibit chaotic behavior in the
superradiant regime \cite{Emary2003,Emary:Chaos}.  In view of this
difference, it would be interesting to explore the possibility of
strange attractors for intermediate $\kappa$. However, for the
parameters we have explored numerically we see no evidence for strange
attractors. Indeed for $g=g^\prime$ and $U>0$ we have demonstrated the
existence of a limit cycle governed by Eq.~(\ref{eq:13}).
Nonetheless, it is worth noting that the nonlinear equations of motion
in Eq.~(\ref{eqmo}) are closely related to the Maxwell--Bloch
equations for a laser \cite{Haken:RMP}.  These are known to be
equivalent to the paradigmatic Lorenz equations \cite{haken75}, the
archetypal example of dissipative chaos.  However, an important
difference from the Maxwell--Bloch equations is the absence of
external driving in Eq.~(\ref{eqmo}). It would be instructive to
explore the ramifications of this in more detail. For work on chaos in
a closely related optomechanical system see
Ref.~\cite{Larson:Photonic}. Further discussion of the phase diagram
for $g \neq g^\prime$ is given in
Appendix~\ref{sec:further-cuts-through}.

\section{Beyond the Effective Dicke Model and its Semiclassical
Treatment}
\label{sec:discussion}

In the preceding sections, we have discussed the semiclassical
dynamics of the nonequilibrium Dicke model, and its relation to
experiments on the self-organization of BECs in optical cavities
\cite{Baumann:Dicke}.  Within the semiclassical description of this
model we have found a rich variety of stable attractors including
non-trivial steady states, persistent oscillations and regimes of
bistability. Having established a wide variety of predictions for the
semiclassical behavior of the open Dicke model, we now consider what
effects may arise in going beyond this effective description. In
Sec.~\ref{sec:effects-beyond-semic} we first consider modifications to
the Dicke model itself, arising from higher momentum states and other
terms in the effective Hamiltonian. In
Sec.~\ref{sec:effects-due-quantum} we briefly comment on the possible
modifications due to higher order quantum effects.

\subsection{Modifications of the Effective Dicke Model}
\label{sec:effects-beyond-semic}

As outlined in Sec.~\ref{Sect:Exp}, the derivation of the effective
Dicke model involves a projection onto the subspace of the two lowest
lying momentum states; see Appendix \ref{sec:depart-from-gener}.
Without this projection, there would also exist coupling to higher
momentum states such as
\begin{equation}
\frac{1}{\sqrt{2}}\sum_{\alpha=\pm} |\alpha 2 k, 0\rangle,\quad
\frac{1}{\sqrt{2}}\sum_{\alpha=\pm} |0, \alpha 2k\rangle, \quad
\frac{1}{2}\sum_{\alpha\beta=\pm} |\alpha 2k, \beta 2k\rangle.
\end{equation}
In general, the occupation of these excited states is expected to be
small for low intensity cavity light fields, as supported by the time
of flight images of Ref.~\cite{Baumann:Dicke}. Nonetheless, there are
regimes of parameter space where these states may be important. In
particular, these high momentum states may destabilize certain phases
predicted by the reduced Dicke model. Specifically, the inverted state
involves excitation to the north pole of the Bloch sphere and may be
susceptible to destabilization.  Indeed, in the parameter regimes
where the effective Dicke model predicts the inverted state, kinetic
approaches predict heating \cite{Niedenzu2011a}. Likewise, regions of
the superradiant phase in which the majority of the atoms are in the
non-zero momentum state may be unstable.

Although the stability of these particular states may be modified, we
anticipate that many of our predictions are only weakly affected by
these additional states. This is supported by the clear quantitative
agreement between the experimentally observed onset of superradiance
and the reduced Dicke model \cite{Baumann:Dicke}.  Our findings within
the projected subspace may also describe experiments exploiting
internal hyperfine states \cite{Dimer:Proposed}, where no higher
levels exist.

In addition to the effects of higher momentum states, one should also
note that the intensities of the forward and backward propagating pump
beams are not of equal magnitude in the experiments of
Ref.~\cite{Baumann:Dicke}; they differ by a factor of $0.6$ due to
losses on reflection. This introduces a coupling to the state
$\tfrac{1}{2}\sum_{\alpha,\beta=\pm} \beta |\alpha k, \beta k\rangle$
which exhibits odd parity under reflection in the pump direction. We
expect such contributions to play a similar role to higher momentum
states.

A further source of possible departure from the idealized Dicke model
with $g=g^\prime$ arises because of the finite atomic recoil
energy. As we demonstrate in Appendix~\ref{sec:corrections-due-non},
the processes leading to $g$ and $g^\prime$ correspond to different
detunings of the intermediate states, so that $\delta g/\bar{g} =
\omega_r (\omega_c-\omega_p) / 2(\omega_a-\omega_p)$.  We have
investigated the possible impact of $g\neq g^\prime$ in
Sec.~\ref{sec:behaviour-g-neq} and in Appendix
\ref{sec:further-cuts-through}, where we showed that differences
$\delta g/\bar{g}$ of the order of $10^{-3}$ can cause one to cross a
phase boundary.  However, for the typical values of the detunings used
in Ref.~\cite{Baumann:Dicke}, this asymmetry is of order $\delta g /
\bar{g} \sim 10^{-12}$. It is therefore too small to have any
significant effect on our findings. Nonetheless, in experiments with a
smaller atom-pump detuning, this asymmetry could play a crucial role.

\subsection{Corrections to Semiclassical Dynamics}
\label{sec:effects-due-quantum}

In the above discussion we have focused primarily on the semiclassical
dynamics of the effective Dicke model through the solutions of the
equations of motion given in Eq.~(\ref{eqmo}). In some places we have
also incorporated the leading effect of quantum fluctuations by using
Wigner distributed initial conditions. This may be interpreted as
including subleading $1/N$ corrections where $N$ is the number of
atoms \cite{Blakie:Dynamics}. More generally it would be profitable to
investigate the full quantum dynamics governed by the density matrix
equations of motion in Eq.~(\ref{eq:1}).  However, it is important to
bear in mind that since the density matrix describes an ensemble
average it will in general mask the effects of spontaneous symmetry
breaking. It will also wash out collective and persistent oscillations
\cite{Milburn:Dynamics}. Nonetheless, in a single experimental run one
still observes spontaneous symmetry breaking
\cite{Baumann:Dicke,Baumann:Dicke2}, and the density matrix describes
the average over many runs; see for example
Ref.~\cite{Leggett:RMPalkali}. In order to recover information on
these non-trivial features within the density matrix formulation, one
may consider higher order correlation functions.  We leave this
problem for future work.

\section{Conclusions}
\label{sec:conclusions}

In this manuscript we have explored the collective dynamics of ultra
cold atoms in transversely pumped optical cavities. Within the
framework of the effective nonequilibrium Dicke model we present a
detailed discussion of the rich phase diagram of asymptotic
attractors, including steady states, coexistence phases and regimes of
persistent oscillations. We show that the inherent timescales for the
destablization of the initial state, and the decay time towards the
asymptotic attractors, show strong variations throughout the dynamical
phase diagram. Crucially, we have demonstrated that two distinct
principal timescales emerge, corresponding to the energy scales
$\omega_0$ and $\omega_0^2/\kappa$. The scale $\omega_0$,
characterizes both the typical frequency of collective oscillations
and their decay rate for a broad range of parameters. The slower scale
$\omega_0^2/\kappa$, governs the decay rate in proximity to dynamical
phase boundaries, and may be interpreted as critical slowing
down. Most notably, in the regime $\omega < \omega_u$, sweep
experiments over $200$ms may be required in order to reach the
asymptotic regime. It would be profitable to explore this
experimentally and we discuss the broad implications for finite
duration experiments.  In particular the superradiant phase divides
into two distinct regimes, denoted SRA and SRB, with the relaxation
rates $\omega_0^2/\kappa$ and $\omega_0$ respectively.  From a
theoretical perspective it would be valuable to investigate the role
of quantum fluctuations, and the effects of states outside of the
two-level Dicke model description. It would also be interesting to
explore the ramifications of these findings in other realizations of
the Dicke model.

\begin{acknowledgments}
  We are very grateful to K. Baumann, F. Brennecke and T. Esslinger
  for helpful discussions of the details of their experiments.  We
  also thank E. Demler, G. Milburn and S. Sachdev for discussions. MJB
  and BDS acknowledge EPSRC grants no. EP/E018130/1 and no.
  EP/F032773/1. JK acknowledges EP/G004714/2 and EP/I031014/1.  This
  research was supported in part by the National Science Foundation
  under Grant No. NSF PHY05-51164.  MJB acknowledges KITP Santa
  Barbara for hospitality during the final stages of this work.
\end{acknowledgments}

\appendix

\section{Derivation of the Generalized Dicke Hamiltonian}
\label{sec:depart-from-gener}

Here we provide a derivation of the generalized Dicke Hamiltonian for
ultra cold atoms placed in an optical cavity, as illustrated in
Fig.~\ref{Fig:selforg}. For simplicity, we consider a homogeneous
system, and neglect the effects of the finite beam waist of the pump
and cavity fields:
\begin{align}
  \label{eq:derivation-starting-point}
  H &= \omega_c \psi^\dagger \psi + \sum_i \left(
    \omega_a \sigma_i^{ee} - \frac{\hbar^2 \nabla_i^2}{2m}
  \right)
  \nonumber\\
  &+
  \sum_i (\sigma_i^{eg} + \sigma_i^{ge}) 
  \biggl[
  g_0 (\psi + \psi^\dagger) \cos(kx_i) 
  \nonumber\\
  &\qquad+ 
  \Omega_f \cos(k z_i - \omega_p t)
  +
  \Omega_b \cos(k z_i + \omega_p t)
  \biggr].
\end{align}
The Hamiltonian acts on both the centre of mass position of the atoms,
and their electronic state.  The latter is restricted to the two
states involved in the optical transitions, denoted by $e$ and $g$ for
the excited and ground states respectively.  In this basis
$\sigma_i^{ee}= |e\rangle_i \langle e|_i$ and $\sigma_i^{eg}=
|e\rangle_i \langle g|_i$. The sum over $i$ runs over the number of
atoms present in the cavity, $N$.  The matter-light interactions
correspond to the dipole coupling of the atomic transition to the
fields of the cavity and the forward and backward pump fields. The
cavity field $\psi$ has explicit quantum dynamics, while the time
dependence of the pump fields is externally imposed.  The cavity-atom
coupling is designated $g_0$.  The pump strengths $\Omega_{f(b)}$
describe the pump beam in the forward (backward) direction, where we
allow for imperfect retro-reflection as discussed in
Ref.~\cite{Baumann:Dicke} and Sec.~\ref{sec:discussion}.  We have
neglected any difference between pump and cavity wavevectors.

The matter-light coupling in Eq.~(\ref{eq:derivation-starting-point})
contains both co- and counter-rotating terms.  The rotating wave
approximation consists of neglecting the counter-rotating terms on the
basis that the detuning is large compared to the coupling strengths.
This approximation is valid here, since $\omega_c, \omega_a, \omega_p$
are all optical frequencies, and of order $400$THz. This is to be
compared to the coupling strengths $\Omega_{f,b} \sqrt{N} \sim g_0
\sqrt{N} \sim 1$ GHz.  Working in the rotating frame at the pump
frequency $\omega_p$, and neglecting the counter-rotating terms, the
Hamiltonian (\ref{eq:derivation-starting-point}) can be rewritten in
the form $H=H_0+H_1$ where
\begin{equation}
   {H}_0 =
   \Delta_c \psi^\dagger \psi
   +  \Delta_a \sum_i  \sigma_i^{ee}
   - \frac{\hbar^2}{2m} \sum_i \nabla_i^2
\end{equation}
and
\begin{equation}
   {H}_1 =
   \sum_i
   \left[
      g_0 {\psi} \cos(kx_i)
      +
      \frac{\Omega_f}{2} e^{i k z_i}  + \frac{\Omega_b}{2} e^{-i k z_i}
     \right] \sigma^{eg}_i + \text{H.c.} .
\end{equation}
Here $\Delta_c=\omega_c-\omega_p$ is the cavity-pump detuning and 
$\Delta_a=\omega_a - \omega_p$ is the atom-pump detuning.

As described in Sec.~\ref{Sect:Exp}, the effective Dicke model is a
low energy description within the electronic ground state, valid if
the coupling to the electronic excited state is small compared to the
detuning $\Delta_a$.  As such, one may proceed by making a
Schrieffer-Wolff~\cite{schrieffer66} transformation and eliminating
the excited electronic state.  This gives a transformed Hamiltonian
${\tilde{H}} = {H}_0 + (i/2)[{S},{H}_1]$ where $[{S},{H}_0]=i {H}_1$.
One should choose ${S}$ so that ${\tilde{H}}$ has no coupling between
the resulting dressed electronic states:
\begin{displaymath}
  i {S} =  \frac{1}{2} \left[
    \Omega_f e^{iz} \hat{f} + \Omega_b e^{-iz} \hat{f}^\ast
    + g_0 {\psi} (e^{ix} \hat{g} + e^{-ix} \hat{g}^\ast) 
  \right] \sigma^{eg} - \text{H.c.},
\end{displaymath}
where for simplicity, position and momenta are now expressed in units
of the cavity wavelength.  In addition, we have suppressed the atom
labels and the summation.  Here the differential operators $\hat{f}$
and $\hat{g}$ are given by
\begin{displaymath}
  \hat{f} = \frac{1}{\Delta_a + \omega_r(1-2 i \partial_z)}, \quad
  \hat{g} = \frac{1}{\Delta_a -\Delta_c + \omega_r(1-2 i \partial_x)},
\end{displaymath}
where $\omega_r = \hbar^2 k^2/2m$ is the recoil energy.
The resulting Hamiltonian has the
form
\begin{align}
  \label{eq:sw-full-H}
  {\tilde{H}} &= {H}_0  - 
  \frac{\Omega_f^2 }{4} \hat{f} - \frac{\Omega_b^2}{4} \hat{f}^\ast
  - \frac{\Omega_f \Omega_b}{8}
  \left(\hat{f} e^{-2iz} + \hat{f}^\ast e^{2iz} + \text{H.c.} \right)
  \nonumber\\ &
  - \frac{g_0^2}{8} {\psi}^\dagger {\psi}
  \left[ \hat{g}(1+e^{-2ix}) + \hat{g}^\ast(1+e^{2ix}) + \text{H.c.} \right]
  \nonumber\\ &
  -  \frac{g_0}{8}
  {\psi}^\dagger (\Omega_f e^{iz} + \Omega_b e^{-iz})
  (\hat{g} e^{-ix} + \hat{g}^\ast e^{ix})
  + \text{H.c.}
  \nonumber\\ &
  -  \frac{g_0}{8}
  {\psi} (e^{ix} + e^{-ix}) 
  (\Omega_f \hat{f} e^{-iz} + \Omega_b \hat{f}^\ast e^{iz})
  + \text{H.c.},
\end{align}
where $\hat{f}^\ast$ and $\hat{g}^\ast$ denote the \emph{complex}
conjugates of the Hermitian operators $\hat{f}$ and $\hat{g}$.  In
writing Eq.~(\ref{eq:sw-full-H}) we have eliminated the excited
electronic states, but have made no further approximations.  As a
result, the form of $\tilde H$ is rather unwieldy. In order to expose
the resulting behavior we will consider two classes of approximation:
the small recoil approximation $\omega_r/\Delta_a \ll 1$ and the weak
pump approximation $q = \Omega_f \Omega_b/(4 \omega_r \Delta_a) \ll
1$.  Both of these approximations are valid for the experiments of
Ref.~\cite{Baumann:Dicke} and we will now consider each in turn.

\subsection{Small Recoil Approximation $\omega_r \ll \Delta_a$}
\label{sec:small-reco-appr}

If the recoil energy $\omega_r$ is small compared to the atom-pump
detuning $\Delta_a$, then one may set $\omega_r=0$ as a first
approximation.  In this case $\hat{f} \simeq \Delta_a^{-1}$ and
$\hat{g} \simeq (\Delta_a-\Delta_c)^{-1}$ become c-numbers and the
form of Eq.~(\ref{eq:sw-full-H}) simplifies considerably.  This
approximation is well justified for the parameters of
Ref.~\cite{Baumann:Dicke} as $\omega_r/\Delta_a \simeq 50 \text{kHz}/
1 \text{THz} \sim 5 \times 10^{-8}$.  In this approximation
Eq.~(\ref{eq:sw-full-H}) becomes
\begin{align}
  \label{eq:sw-small-recoil}
  {\tilde{H}}&=
  -\omega_r \nabla^2
  -
  \frac{1}{4\Delta_a} \left[
    \Omega_f^2 + \Omega_b^2 + 2  \Omega_f \Omega_b \cos(2z)
  \right]
  \nonumber\\
  &+\left[ \Delta_c
    - \frac{g_0^2}{\Delta_a-\Delta_c}
    \cos^2(x)
  \right] \psi^\dagger \psi
  \nonumber\\
  &- \frac{g_0}{4}
  \left[ \frac{1}{\Delta_a} + \frac{1}{\Delta_a - \Delta_c}\right]
  \cos(x)
  \nonumber \\
  &\times \Bigl[
  i(\Omega_f - \Omega_b) \sin(z)  (\psi^\dagger - \psi)
  \nonumber\\&\quad+
  (\Omega_f + \Omega_b) \cos(z) (\psi^\dagger + \psi)
  \Bigr].
\end{align}
We proceed by introducing a basis set of atomic center of mass states
for the atoms in their electronic ground state. The first two of these
states will correspond to the ``spin down'' and ``spin up'' states of
the effective Dicke model.  These basis states are given by the
eigenstates of the first line of Eq.~(\ref{eq:sw-small-recoil}), which
may be written as
\begin{equation}
  \label{eq:basis}
  \Phi_{\sigma,m,n} (x,z) =  \phi_{\sigma,m}(z)
  \begin{cases}
    \frac{\cos(nx)}{\sqrt{\pi}}; & n>0, \\
    \frac{1}{\sqrt{2\pi}}; & n =0.
  \end{cases}
\end{equation}
The  energies are given by 
$E = \omega_r (n + \varepsilon_{\sigma,m}) + \text{const.}$
where
\begin{equation}
  \label{eq:mathieu}
  \frac{d^2 \phi_{\sigma,m}}{d z^2}  + \left[ \varepsilon_{\sigma,m}
    - 2 q \cos(2 z) \right]
  \phi_{\sigma,m} = 0
\end{equation}
is the Mathieu equation \cite{Whittaker:Modern} and $\sigma=\pm$ label
the even and odd solutions. The Mathieu parameter $q = \Omega_f
\Omega_b/(4 \Delta_a \omega_r)$ is a dimensionless measure of the
pumping strength. Due to the form of the matrix elements arising from
the terms linearly dependent on $g_0$ in
Eq.~(\ref{eq:sw-small-recoil}), not all of the configurations of
$\sigma, m, n$ are coupled.  Only those states that can be reached by
a sequence of absorption and emission processes, starting from an atom
in the ground state, need to be included.  If $\Omega_b=\Omega_f$,
then only the even Mathieu functions $\phi_{+,m}(z)$ need to be
included. However, if $\Omega_b \neq \Omega_f$, the odd Mathieu
functions $\phi_{-,m}(z)$ should also be considered.

To recover the effective Dicke model one must restrict attention to
the two lowest states, and work in the limit
$\Omega_f=\Omega_b=\Omega$. In this case, the lowest coupled states
are $\Phi_{+,0,0}$ and $\Phi_{+,1,1}$.  The values of the parameters
$\omega,\omega_0,U$ in the effective Dicke model in
Eq.~(\ref{Dickeham}) can be found by evaluating $\langle \Phi_{+,0,0}
|{\tilde{H}} | \Phi_{+,0,0} \rangle$ and $\langle \Phi_{+,1,1}
|{\tilde{H}} | \Phi_{+,1,1} \rangle$. In terms of the Dicke model
parameters in Eq.~(\ref{Dickeham}), the energies of a configuration
with $n_{\text{ph}}$ photons and all $N$ atoms in either their ground
or excited states are given by:
\begin{equation}
  \label{eq:energies-from-dicke-coeff}
  E_{\downarrow,\uparrow}
  = \mp \frac{\omega_0}{2} N
  + \left(\omega \mp \frac{UN}{2}\right)  n_{\text{ph}}.
\end{equation}
By comparing $E_\downarrow$ and $E_\uparrow$ with the expressions for
$\langle \Phi_{+,0,0} |{\tilde{H}} | \Phi_{+,0,0} \rangle$ and
$\langle \Phi_{+,1,1} |{\tilde{H}} | \Phi_{+,1,1} \rangle$ one may
identify the coefficients $\omega$, $\omega_0$ and $U$.  We find
\begin{equation}
  \label{eq:defn-cav-params}
  \omega = \Delta_c - \frac{5 g_0^2 N}{8 (\Delta_a - \Delta_c)}
  \quad
  U=- \frac{g_0^2}{4 (\Delta_a - \Delta_c)},
\end{equation}
where we have carried out the summation over atoms and made use of the
results $\langle \Phi_{+,0,0} |\cos^2(x)|\Phi_{+,0,0} \rangle = 1/2$
and $\langle \Phi_{+,1,1}|\cos^2(x)| \Phi_{+,1,1}\rangle = 3/4$.
These coefficients agree with those of Ref.~\cite{Nagy:Dicke} when the
pump and cavity frequencies are near detuned. The two-level energy
splitting is given by the difference of the eigenvalues of the states
written in Eq.~(\ref{eq:basis}) and so
\begin{equation}
  \label{eq:defn-atom-params}
  \omega_0 = \omega_r( 1 + \varepsilon_{+,1} - \varepsilon_{+,0}).
\end{equation}
Evaluating the off-diagonal elements $\langle \Phi_{+,0,0} |
{\tilde{H}} | \Phi_{+,1,1} \rangle$ and equating $\langle \Phi_{+,0,0}
|{\tilde{H}} | \Phi_{+,1,1} \rangle = g {\psi}^\dagger + g^\prime
{\psi}$ one finds the remaining Dicke model parameters
\begin{multline}
  \label{eq:defn-g}
    g = g^\prime = 
    -  \frac{g_0 \Omega }{2}
    \left[\frac{1}{\Delta_a} + \frac{1}{\Delta_a-\Delta_c}
    \right]\times
    \\
    \frac{1}{\sqrt{2}} \int_{-\pi}^\pi  dz\, \phi_{+,0}(z) \cos(z) \phi_{+,1}(z).
\end{multline}

Up until this point all the results we have derived in
Eqs.~(\ref{eq:defn-cav-params}) -- (\ref{eq:defn-g}) are formally
exact for arbitrary $q= \Omega_f \Omega_b/(4 \Delta_a \omega_r)$.
However, the fact that one can restrict to the two lowest momentum
states is only valid for weak pumping, i.e. small $q$.  We will
therefore focus on the small $q$ limit; in
Sec.~\ref{sec:equat-moti-extend} we will return to the general $q$
case, including also the presence of higher momentum states.  In the
small $q$ limit one obtains
\begin{equation}
  \label{eq:defn-g-small-q}
  \omega_0\approx 2 \omega_r,
  \quad
  g = g^\prime \approx
  -  \frac{g_0 \Omega }{4}
  \left(\frac{1}{\Delta_a} + \frac{1}{\Delta_a-\Delta_c} \right),
\end{equation}
where we have used the approximations $\phi_{+,0}(z) \approx
1/\sqrt{2\pi}$ and $\phi_{+,1} \approx \cos(z)/\sqrt{\pi}$.  If one
further neglects the cavity-pump detuning $\Delta_c$ in comparison to
the atom-pump detuning $\Delta_a$, then these expressions reduce to
those given in Ref.~\cite{Baumann:Dicke}.
In Sec.~\ref{sec:corrections-due-non} we will generalize the results
of this section to include the effects of nonzero $\omega_r/\Delta_a$.

\subsection{Corrections Due to Non-Zero $\omega_r/\Delta_a$}
\label{sec:corrections-due-non}

In order to quantify the effects of non-zero $\omega_r/\Delta_a$, we
return to Eq.~(\ref{eq:sw-full-H}), but continue to make use of the
$q=0$ approximation used in the second half of the previous section.
The main difference that non-zero $\omega_r$ introduces is that $g$
and $g^\prime$ are no longer equal.  One obtains
\begin{equation}
  \label{eq:finomr-g}
  \begin{aligned}
  g&=  - \frac{g_0 \Omega}{4} 
  \left[
    \frac{1}{\Delta_a+\omega_r-\Delta_c}
    +
    \frac{1}{\Delta_a-\omega_r}
  \right],
  \\
  g^\prime &=
  - \frac{g_0 \Omega}{4} 
   \left[ 
     \frac{1}{\Delta_a-\omega_r-\Delta_c}
     +
     \frac{1}{\Delta_a+\omega_r}
  \right], 
  \end{aligned}
\end{equation}
where for simplicity we set $\Omega_f=\Omega_b=\Omega$.  In the limit
$\omega_r \to 0$, these reduce to Eq.~(\ref{eq:defn-g}).  In
Sec.~\ref{sec:behaviour-g-neq}, we show that a phase boundary can be
crossed if the fractional difference $\delta g/\bar{g}$ is large
enough.  At leading order in $\omega_r/\Delta_a$ the fractional
difference given by Eq.~(\ref{eq:finomr-g}) is $\delta g/\bar{g} =
\omega_r \Delta_c / \Delta_a^2$. For the experimental parameters in
Ref.~\cite{Baumann:Dicke} this fractional difference is too small to
cross the phase boundary.  However, for smaller $\Delta_a$ this
fractional difference may become significant.

The results in Eq.~(\ref{eq:finomr-g}) are obtained by the same
procedure as in the previous section, by evaluating the off-diagonal
matrix element and equating $\langle \Phi_{+,0,0} |{\tilde{H}} |
\Phi_{+,1,1} \rangle = g {\psi}^\dagger + g^\prime {\psi}$. 
In deriving Eq.~(\ref{eq:finomr-g}) we use the basis states
$\Phi_{+,0,0}(x,z) = 1/(2\pi), \Phi_{+,1,1}(x,z) =
\cos(x)\cos(z)/\pi$, which follow from Eq.~(\ref{eq:mathieu}) at
$q=0$.  We further employ the identities
\begin{align*}
  \int_{-\pi}^{\pi} \frac{d \zeta}{2\pi} \left[
    \hat{\mathcal{D}}
    e^{-i\zeta}
    +
    \hat{\mathcal{D}}^\ast
    e^{i\zeta}
  \right] \cos(\zeta)
  &= \frac{1}{x},
  \\
    \int_{-\pi}^{\pi} \frac{d \zeta}{2\pi} \left[
    e^{i\zeta} 
    \hat{\mathcal{D}}
    +
    e^{-i\zeta}     \hat{\mathcal{D}}^\ast
  \right] \cos(\zeta)
  &= \frac{1}{x-2 \omega_r},
\end{align*}
where $ \hat{\mathcal{D}} = [x - i 2 \omega_r \partial_\zeta]^{-1}$.

The remaining parameters of the Dicke model are
\newcommand{\dshift}{\delta}
\begin{align*}
    U&= -
    \frac{g_0^2}{4}
  \left[
    \frac{1}{ \dshift + 3\omega_r }
    + 
    \frac{2}{ \dshift - \omega_r }
    -
    \frac{2}{ \dshift + \omega_r }
  \right]    
  \\%  \label{eq:finite-omr-om}
  \omega &= \Delta_c - \frac{g_0^2}{8}
  \left[
    \frac{1}{\dshift + 3\omega_r }
    +
    \frac{2}{\dshift - \omega_r }
    +
    \frac{2}{\dshift + \omega_r }
  \right]    
  \\%  \label{eq:finite-omr-om0}
  \omega_0 &= 2 \omega_r 
  - \frac{\Omega^2}{4} \left[
    \frac{1}{\Delta_a + 3 \omega_r}
    +
    \frac{2}{\Delta_a - \omega_r}
    -
    \frac{2}{\Delta_a+\omega_r}
  \right]
\end{align*}
where $\dshift = \Delta_a - \Delta_c \equiv \omega_a - \omega_c$.  In
the limit of $\omega_r \to 0$ these reduce to
Eq.~(\ref{eq:defn-cav-params}) and the small $q$ expansion of
Eq.~(\ref{eq:defn-atom-params}).  These expressions are found using
the same procedure as outlined in the previous section, by equating
$E_\downarrow,E_\uparrow$ in Eq.~(\ref{eq:energies-from-dicke-coeff})
with the expressions for $\langle \Phi_{+,0,0} |{\tilde{H}} |
\Phi_{+,0,0} \rangle$ and $\langle \Phi_{+,1,1} |{\tilde{H}} |
\Phi_{+,1,1} \rangle$.  To evaluate these expressions we use the
results $\hat{f} |\Phi_{+,0,0}\rangle = (\Delta_a + \omega_r)^{-1}
|\Phi_{+,0,0}\rangle$ and $\hat{g} |\Phi_{+,0,0}\rangle = (\Delta_a +
\omega_r -\Delta_c)^{-1}|\Phi_{+,0,0}\rangle$ together with
\begin{multline*}
  \int_{-\pi}^{\pi} \frac{d\zeta}{\pi} \cos(\zeta) \left[
    \frac{1}{x-i 2\omega_r \partial_\zeta} ( 1 + e^{-2i\zeta}) + \text{c.c.} +
    \right.\\\left.
      ( 1 + e^{2i\zeta}) \frac{1}{x-i 2\omega_r \partial_\zeta} + \text{c.c.}
  \right] \cos(\zeta)
  = \frac{4}{x-2\omega_r} + \frac{2}{x+2\omega_r}
\end{multline*}
where $x=\Delta_a+\omega_r$ and $x=\Delta_a+\omega_r-\Delta_c$ for integrals involving
$\hat{f}$ and $\hat{g}$ respectively.  

\subsection{Equations of motion in extended state space}
\label{sec:equat-moti-extend}

When the parameter $q$ is not small, the restriction to two atomic
states is no longer valid.  Including higher momentum states, it is no
longer possible to map the problem on to an effective spin
Hamiltonian. However, it is still possible to derive a semiclassical
description of the coupled atom and cavity system.  In this
generalised case, the semiclassical description consists of a
Gross-Pitaevskii equation for a macroscopically occupied atomic
wavefunction $\chi(x,z)$ coupled to a quasi-classical Heisenberg
equation for the photon field $\psi$.

Decomposing the atomic wavefunction in some basis $\chi(x,z) =
\sum_\alpha \chi_\alpha \Phi_\alpha(x,z)$ and splitting the Hamiltonian
into the parts
\begin{displaymath}
  {\tilde{H}} =  {h}^{(0)} + 
  {\psi}^\dagger {\psi}   {h}^{(1)} 
  +
  ({\psi}^\dagger+{\psi}) {h}^{(2)}
  +
  i({\psi}^\dagger-{\psi}) {h}^{(3)}
\end{displaymath}
the explicit equations of motion read
\begin{align}
  \label{eq:eom-ext}
  i \partial_t \chi_\alpha &= 
  \left(
     M_{\alpha\beta}^{(0)} + 
     |\psi|^2  M_{\alpha\beta}^{(1)}
  \right.\nonumber\\&\left.\qquad+
     (\psi^\ast+\psi) M_{\alpha\beta}^{(2)}
    +
     i(\psi^\ast-\psi) M_{\alpha\beta}^{(3)}
  \right) \chi_\beta,
  \\
  i \partial_t \psi &= \left( 
    \chi^\ast_\alpha M_{\alpha\beta}^{(1)} \chi_\beta
    - i \kappa\right) \psi 
  \nonumber\\ &\qquad
  +  \chi^\ast_\alpha M_{\alpha\beta}^{(2)} \chi_\beta
  + i  \chi^\ast_\alpha M_{\alpha\beta}^{(3)} \chi_\beta,
\end{align}
where have defined $M^{(n)}_{\alpha\beta} \equiv \langle \Phi_\alpha
|{h}^{(n)}| \Phi_\beta \rangle$.  As long as $q$ is small, one may
truncate to two atomic basis states, and thereby recover the
semiclassical equations in Eq.~(\ref{eqmo}), where $S_z = |\chi_1|^2 -
|\chi_0|^2$ and $S_x + i S_y = \chi_1^\ast \chi_0$.

\section{Fixed Points with Arbitrary  $g$ and $g^\prime$}
\label{App:Genfp}
In general it is difficult to obtain explicit closed form expressions
for the steady state solutions of Eq.~(\ref{eqmo}). However, in the
special case where $U=0$ simplifications occur for arbitrary $g$ and
$g^\prime$. More generally, for arbitrary $U$, $g$ and $g^\prime$ one
may obtain self-consistent implicit solutions. We
discuss these cases below.

\subsection{$U=0$}

In the case where $U=0$ the nonlinear equations become linear
equations for the variables $\psi_1$, $\psi_2$, $S_x$ and $S_y$, where
$S_z$ enters via the coefficients.  Decomposing Eqs.~(\ref{eqmo1}) and
(\ref{eqmo2}) into their real and imaginary parts yields
\begin{equation}
\begin{aligned}
\omega_0 S_x & = 2(g+g^\prime)S_z\psi_1,\\
-\omega_0 S_y & = 2(g-g^\prime) S_z\psi_2,\\
\kappa\psi_1-\omega\psi_2 & =- (g-g^\prime)S_y, \\
\omega\psi_1+\kappa\psi_2 & =-(g+g^\prime)S_x.
\end{aligned}
\label{ssuzero}
\end{equation}
The last two equations may also be written in the form
\begin{equation}
\begin{aligned}
(\omega^2+\kappa^2) \psi_1 & = -\omega(g+g^\prime)S_x-\kappa (g-g^\prime)S_y, \\
(\omega^2+\kappa^2)\psi_2 & = -\kappa(g+g^\prime)S_x+\omega(g-g^\prime)S_y.
\end{aligned}
\end{equation}
The condition for
non-trivial solutions yields the determinantal self-consistency
equation
\begin{equation}
4(g^2-{g^\prime}^2)^2 S_z^2+4\omega\omega_0(g^2+{g^\prime}^2)S_z+(\omega^2+\kappa^2)\omega_0^2=0.
\label{szquaduzero}
\end{equation}
This may be solved to yield
\begin{equation}
S_z=\frac{-\omega\omega_0(g^2+{g^\prime}^2)\pm 
\sqrt{(2\omega\omega_0 g g^\prime)^2-\omega_0^2\kappa^2(g^2-{g^\prime}^2)^2}}
{2(g^2-{g^\prime}^2)^2},
\label{szu0}
\end{equation}
corresponding to a non-trivial superradiant phase with $\psi\neq 0$.
In the limit $g=g^\prime$ one obtains
$S_z=-\omega_0(\omega^2+\kappa^2)/8\omega g^2$. The critical coupling
strength for the onset of superradiance corresponds to $S_z=-N/2$ and
is given by
\begin{equation}
g\sqrt{N}=\sqrt{\frac{\omega_0(\omega^2+\kappa^2)}{4\omega}},
\label{appgsra}
\end{equation} 
in agreement with the results of Dimer {\em et al}
\cite{Dimer:Proposed}. It also coincides with Eq.~(\ref{sragc}) for
the onset of the SRA phase when $U=0$. The explicit dependence on
$\kappa$ in Eq.~(\ref{appgsra}) emphasizes that the transition occurs
in an open system.  In the limit $\kappa=0$ one recovers the location
of the superradiance transition, $g\sqrt{N}=\sqrt{\omega\omega_0}/2$,
for the equilibrium Dicke model with counter-rotating terms.

\subsection{$U\neq 0$}
In the general case with arbitrary $U$, $g$ and $g^\prime$ the steady state solutions of Eq.~(\ref{eqmo}) are more difficult
to obtain in an explicit form.  
However, one may still obtain self-consistent solutions
which relate the photon density to $S_z$ for example. In turn, these
implicit consistency equations may be solved numerically. For generic
parameters the steady state equations of motion may be obtained from
Eqs.~(\ref{ssuzero}) by replacing $\omega_0\rightarrow\tilde\omega_0$
and $\omega\rightarrow\tilde\omega$, where $\tilde\omega\equiv \omega
+U S_z$, $\tilde\omega_0\equiv\omega_0+Un$ and $n\equiv
|\psi|^2=\psi_1^2+\psi_2^2$. For a given photon occupation number,
$n$, non-trivial solutions satisfy the determinantal
Eq.~(\ref{szquaduzero}) with these replacements. Using the explicit
form of $\tilde\omega\equiv \omega+US_z$ one obtains a modified
quadratic equation for $S_z$:
\begin{equation}
  (\chi^2-16g^2{g^\prime}^2)S_z^2+2\omega\tilde\omega_0\chi S_z+(\omega^2+\kappa^2)\tilde\omega_0^2=0,
\label{genquad}
\end{equation}
where we define 
$\chi\equiv 2(g^2+{g^\prime}^2)+U\tilde\omega_0$. This has solutions
\begin{equation}
S_z=\frac{\tilde\omega_0}{\chi^2-16g^2{g^\prime}^2}\left[-\omega\chi\pm 
\sqrt{16g^2{g^\prime}^2(\omega^2+\kappa^2)-\kappa^2\chi^2}\right].
\label{gensz}
\end{equation}
In order to have real solutions to Eq.~(\ref{gensz}) one requires
$16g^2{g^\prime}^2(\omega^2+\kappa^2)-\kappa^2\chi^2\ge 0$. This
translates into the condition that ${\rm max}(0,n_-)\le n\le n_+$
where
\begin{equation}
n_\pm =U^{-2}\left[-2(g^2+{g^\prime}^2)-U\omega_0
\pm 4gg^\prime\kappa^{-1}\sqrt{\omega^2+\kappa^2}\right].
\nonumber
\end{equation}
One should restrict attention to those cases where
$|S_z|<N/2$. Having found a solution for $S_z$ in terms of the number
of photons $n$, one may also find an equation for $n$ in terms of
$S_z$. Using the analogues of the first two equations in
Eq.~(\ref{ssuzero}) with $\omega_0\rightarrow\tilde\omega_0$
\begin{equation}
  |\psi|^2=\psi_1^2+\psi_2^2=\frac{\tilde\omega_0^2S_x^2}{S_z^2}\left[\frac{1}{4(g+g^\prime)^2}+\frac{r^2}{4(g-g^\prime)^2}\right],
\label{n1}
\end{equation}
where $r\equiv S_y/S_x$. Using the fixed length spin constraint ${\bf
  S}^2=N^2/4$ one may eliminate $S_x^2=(N^2/4-S_z^2)/(1+r^2)$ in
favour of the ratio $r$. This ratio may be obtained by using the
analogues of the first two of Eqs.~(\ref{ssuzero}) to subsitute
$\psi_1$ and $\psi_2$ into the analogue of the third equation:
\begin{equation}
r=\mathcal{A} \left(\frac{g-g^\prime}{g+g^\prime}\right); \quad 
\mathcal{A}=-\frac{\tilde\omega_0\kappa}{\tilde\omega\tilde\omega_0+2(g-g^\prime)^2S_z}.
\end{equation}
Substituting into Eq.~(\ref{n1}) yields
\begin{equation}
n=|\psi|^2=\frac{\tilde\omega_0^2}{4S_z^2}\frac{(N^2/4-S_z^2)(1+\mathcal{A}^2)}{(g+g^\prime)^2+(g-g^\prime)^2\mathcal{A}^2}.
\end{equation}

\section{Linear Stability for Arbitrary States}
\label{App:linstab}
In Sec.~\ref{sec:stability} we linearized the equations of motion
around the normal ($\Downarrow$) and inverted ($\Uparrow$) states. The
corresponding self-consistency equation is given by
\begin{equation}
\begin{vmatrix}
\eta +i\kappa-\omega_\mp & 0 & -g & -g^\prime\\
0 & \eta+i\kappa +\omega_\mp  & g^\prime & g \\
\mp gN & \mp g^\prime N & \eta-\omega_0 & 0 \\
\pm g^\prime N & \pm gN & 0 & \eta+\omega_0
\end{vmatrix}=0.
\label{niselfcon}
\end{equation}
This yields a quartic equation
\begin{equation}
\begin{gathered}
  0=\left[(\eta+i\kappa)^2-\omega_\mp^2\right](\eta^2-\omega_0^2)+(g^2N-{g^\prime}^2N)^2\\
  \mp 2\eta(\eta+i\kappa)(g^2N-{g^\prime}^2N)\mp
  2(g^2N+{g^\prime}^2N)\omega_\mp\omega_0,
\label{linquart}
\end{gathered}
\end{equation}
whose roots characterize the possible instabilities. 
In the case where $g=g^\prime$ Eq.~(\ref{linquart}) reduces to
Eq.~(\ref{equalgquart}). More generally one may linearize 
the equations of motion in Eq.~(\ref{eqmo}) around an
arbitrary state, so that $\psi=\psi_0+\delta\psi$, $S^-=S_0^-+\delta
S^-$ and $S^z=S^z_0+\delta S^z$, one obtains
\begin{equation}
\begin{aligned}
  \dot{\delta S^-} = & -i\tilde\omega_0\delta S^--iU(\psi_0^\ast\delta\psi+\psi_0\delta\psi^\ast)S_0^-\\
  & +2i(g\psi_0+g^\prime\psi^\ast_0)\delta
  S^z+2i(g\delta\psi+g^\prime\delta\psi^\ast)S^z_0,\\
\dot{\delta S^z} = & -ig\delta\psi S_0^++ig\delta \psi^\ast S_0^-+ig^\prime\delta\psi S_0^--ig^\prime\delta\psi^\ast S_0^+ \\
& -ig\psi_0\delta S^++ig\psi_0^\ast \delta S^-+ig^\prime \psi_0\delta S^--ig^\prime\psi_0^\ast\delta S^+,\\
\dot{\delta \psi} = & - (\kappa+i\tilde\omega)\delta\psi-iU\psi_0\delta S^z-ig\delta S^--ig^\prime\delta S^+,
\label{fulllin}
\end{aligned}
\end{equation}
where $\tilde\omega=\omega + US^z_0$ and
$\tilde\omega_0=\omega_0+U|\psi_0|^2$.  Parameterizing $\delta\psi=a
e^{-i\eta t}+b^\ast e^{i\eta^\ast t}$, $\delta S^-=ce^{-i\eta t}+d^\ast
e^{i\eta^\ast t}$ and $\delta S^z=fe^{-i\eta t}+f^\ast e^{i\eta^\ast t}$,
one obtains a set of algebraic equations for the coefficients $a$,
$b$, $c$, $d$ and $f$. The corresponding secular equation is given by
$|\eta {\mathbb I} -{\mathbb M}|=0$, where ${\mathbb I}$ is a $5\times
5$ unit matrix and
\begin{widetext}
\begin{equation}
{\mathbb M}=
\begin{pmatrix}
\tilde\omega -i\kappa & 0 & g & g^\prime & U\psi_0 \\
0 & -(\tilde\omega+i\kappa) & -g^\prime  & -g & -U\psi_0^\ast \\
-2gS_0^z+U\psi_0^\ast S_0^- & -2g^\prime S_0^z+U\psi_0S_0^- &\tilde\omega_0 
& 0 & -2(g\psi_0+g^\prime\psi_0^\ast) \\
2g^\prime S_0^z-U\psi_0^\ast S_0^+ & 2gS_0^z-U\psi_0S_0^+ & 0 & -\tilde\omega_0 & 
2(g\psi_0^\ast+g^\prime\psi_0) \\
gS_0^+-g^\prime S_0^- & -gS_0^-+g^\prime S_0^+ & -(g\psi_0^\ast + g^\prime \psi_0) & 
(g\psi_0+g^\prime\psi_0^\ast) & 0
\end{pmatrix}.
\label{m5}
\end{equation}
\end{widetext}
In general, there are in fact only four independent equations owing to
the fixed length constraint, ${\bf S}^2=N^2/4$. This is reflected by a
redundant zero mode, $\eta=0$, corresponding to longitudinal
fluctuations in the length of the spin. Although Eq.~(\ref{m5})
captures all of the essential information, it is convenient to
eliminate this zero mode and use a $4\times 4$ matrix representation
for the physical transverse degrees of freedom. 
Differentiating the fixed length constraint with respect to time
yields $2S^z\dot S^z+(S^+{\dot S}^-+S^-{\dot S}^+)=0$. Linearizing
around a fixed point gives $2S_0^z\dot{\delta S^z}+S_0^+\dot{\delta
  S^-}+S_0^-\dot{\delta S^+}=0$, as may verified by
Eq.~(\ref{fulllin}) and Eq.~(\ref{eqmo}). Using the normal mode
parameterization one obtains the relationship
$f=-(S_0^+c+S_0^-d)/2S_0^z$, between the coefficients. Eliminating $f$
from the linear equations yields $|\eta \tilde{\mathbb
  I}-\tilde{\mathbb M}|=0$, where $\tilde{\mathbb I}$ is a $4\times 4$
unit matrix and
\begin{widetext}
\begin{equation}
\tilde{\mathbb M}=
\begin{pmatrix}
  \tilde\omega-i\kappa & 0 & g-U\psi_0S_0^+/(2S_0^z) &
  g^\prime -U\psi_0S_0^-/(2S_0^z)\\
  0 & -(\tilde\omega+i\kappa)& -g^\prime+U\psi_0^\ast
    S_0^+/(2S_0^z)
  & -g+U\psi_0^\ast S_0^-/(2S_0^z)\\
  -2gS_0^z+U\psi_0^\ast S_0^- & -2g^\prime S_0^z+U\psi_0S_0^- &
  \tilde\omega_0 +(g\psi_0+g^\prime\psi_0^\ast) S_0^+/S_0^z &
  (g\psi_0+g^\prime\psi_0^\ast)S_0^-/S_0^z \\
  2g^\prime S_0^z-U\psi_0^\ast S_0^+ & 2g S_0^z-U\psi_0 S_0^+&
  -(g\psi_0^\ast+g^\prime\psi_0) S_0^+/S_0^z &
  -\tilde\omega_0-(g\psi_0^\ast+g^\prime\psi_0) S_0^-/S_0^z
\end{pmatrix}.
\label{m4}
\end{equation}
\end{widetext}
\subsection{Stability of the normal ($\Downarrow$) and inverted
  ($\Uparrow$) states with arbitrary $g$ and $g^\prime$}
\label{App:stab-norm-down}

In the case of the normal ($\Downarrow$) and inverted ($\Uparrow$)
states where $\psi_0=0$ and $S^z=\mp N/2$, the determinantal equation
$|\eta \tilde{\mathbb I}-\tilde{\mathbb M}|=0$ reduces to
Eq.~(\ref{niselfcon}). This corresponds to the quartic equation given
in Eq.~(\ref{linquart}).  In order to find when the roots become
unstable it is convenient to decompose the quartic polynomial into
its real and imaginary parts so that $A_\mp(\eta)+iB_\mp(\eta)=0$ where
\begin{equation}
\begin{aligned}
A_\mp(\eta) & =(\eta^2-\omega_\mp^2-\kappa^2)(\eta^2-\omega_0^2)+(g^2N-{g^\prime}^2N)^2\\
& \mp 2(g^2N+{g^\prime}^2N)\omega_\mp\omega_0\mp 2\eta^2(g^2N-{g^\prime}^2N),\\
B_\mp(\eta) & =2\kappa\eta\left[\eta^2-\omega_0^2\mp (g^2N-{g^\prime}^2N)\right].
\end{aligned}
\end{equation}
We may thus find the roots of the equation $B_\mp(\eta)=0$, and
subsequently impose the condition $A_\mp(\eta)=0$ on these solutions.
It is readily seen that $B_\mp(\eta)=0$ has a solution $\eta=0$
corresponding to an exponentially growing mode without
oscillations. Substituting this value into the expression for
$A_\mp(\eta)$ yields the condition $A_\mp (0)=0$:
\begin{equation}
(\omega_\mp^2+\kappa^2)\omega_0^2+(g^2N-{g^{\prime}}^2N)^2\mp 2(g^2N+{g^\prime}^2N)\omega_\mp\omega_0=0.
\label{Acond}
\end{equation}
In the special case where $g=g^\prime$ one recovers the critical
condition for the onset of the SRA phase as given by
Eq.~(\ref{sragc}). Alternatively, there is also an instability with a
finite frequency $\eta^2=\omega_0^2\pm(g^2N-{g^\prime}^2N)$
corresponding to $B_\mp(\eta)=0$. Demanding that $A_\mp(\eta)=0$
yields a critical condition for the ratio of the couplings
\begin{equation}
\frac{{g^\prime}^2}{g^2}=\frac{(\omega_\mp+\omega_0)^2+\kappa^2}{(\omega_\mp-\omega_0)^2+\kappa^2},
\label{ggratio}
\end{equation}
rather than their absolute scales. This instability condition
manifests itself as the phase boundaries shown in the upper panels of
Figs.~\ref{fig:omega-g-negative-u-vs-dg} and
\ref{fig:omega-g-positive-u-vs-dg}. Denoting $\delta g\equiv
g^\prime-g$, $\bar g\equiv (g+g^\prime)/2$, and noting that
$\omega_0\ll \kappa$, the phase boundary given by Eq.~(\ref{ggratio})
may be approximated as
\begin{equation}
\frac{\delta g}{\bar g}\approx \frac{2\omega_0\omega_\mp}{\omega_\mp^2+\kappa^2}.
\label{dggbarratio}
\end{equation}
In a similar fashion Eq.~(\ref{Acond}) may be recast as 
\begin{equation}
{\bar g}^2\approx \frac{(g_a^\mp)^2-\delta g^2/4}{1-\delta g^2N/\omega_0\omega_\mp},
\label{ghoriz}
\end{equation}
where $g_a^\mp$ is given by Eq.~(\ref{sragc}). For the range of
parameters shown in Figs.~\ref{fig:omega-g-negative-u-vs-dg} and
\ref{fig:omega-g-positive-u-vs-dg}, $\delta g\ll \bar g, g_a^\mp$, and
Eq.~(\ref{ghoriz}) is effectively independent of $\delta g/\bar g$;
this yields the horizontal phase boundaries in
Figs.~\ref{fig:omega-g-negative-u-vs-dg} and
\ref{fig:omega-g-positive-u-vs-dg}.

\subsection{Stability of SRA and SRB with $g=g^\prime$}
\label{srabstab}
When $g=g^\prime$ the SRA phase has $S^y=0$ and $S^\pm=S^x$. In this case
the matrix $\tilde {\mathbb M}$ has eigenvalues $\eta$, satisfying
\begin{equation}
\begin{aligned}
  \left[(\eta+i\kappa)^2-{\tilde\omega}^2\right]
& \left[\eta^2-({\tilde\omega}_0 N/2 S_0^z)^2\right]\\
  &
  +\frac{2{\tilde\omega}{\tilde\omega}_0}{S^z_0}|2gS_0^z-U\psi_0S_0^x|^2=0,
\end{aligned}
\label{appsrapoly}
\end{equation}
where $\tilde\omega=\omega+US_0^z$ and 
$\tilde\omega_0=\omega_0+U|\psi_0|^2$. 

When $g=g^\prime$ the SRB phase has $\tilde\omega=\tilde\omega_0=0$
and $\psi\equiv \psi_1+i\psi_2$ is purely imaginary. In this case the
diagonal blocks of $\tilde{\mathbb M}$ are proportional to unity and
zero respectively.  The eigenvalues satisfy Eq.~(\ref{srbchar}) which
is the square of a quadratic equation. The exact eigenvalues
corresponding to fluctuations around the stable SRB fixed points are
given by Eq.~(\ref{eq:9}).

\section{Transitions Near the Tricritical Points}
\label{App:Bifurcations}
\begin{figure}[!t]
 \centering
 \includegraphics[width=3.2in]{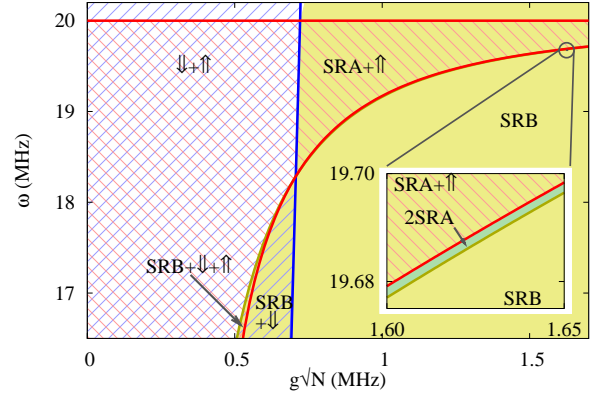}
 \caption{(color online). 
Magnified portion of the bottom panel of
   Fig.~\ref{fig:omega-g-negative-u} in the vicinity of the
   tricritical point where three phase boundaries cross. In addition
   to the phases visible in Fig.~\ref{fig:omega-g-negative-u} there is
   a narrow region denoted as 2SRA, where the two distinct SRA
   solutions given by Eq.~(\ref{szsra}) coexist; see inset.}
\label{trizoom}
\end{figure}

As noted in Ref.~\cite{Keeling:Collective} for $UN<-2\kappa$, three of
the phase boundaries in Fig.~\ref{fig:omega-g-negative-u} cross at the
point $\omega = \sqrt{\omega_u^2 - \kappa^2}, g=\sqrt{-\omega_0
  U/4}$. As shown in Fig.~\ref{trizoom}, in this vicinity there is a
narrow region where the two distinct SRA solutions given by
Eq.~(\ref{szsra}), together with their parity symmetry partners, are
stabilized; see also Fig.~\ref{srasrbtran}.
On Fig.~\ref{fig:omega-g-negative-u} these occur within the width of
the line marking the boundary of the $\text{SRA}+\Uparrow$ and
$\text{SRB}$ regions.  At $g=g_b$ the two pairs of SRA solutions
merge, and switch to two pairs of SRB solutions. After this, one of
each pair is stable whilst the others are unstable, as generically
occurs in the SRB phase.

\section{Further cuts through the phase
  diagram with $g \neq g^\prime$ and $U\neq 0$}
\label{sec:further-cuts-through}
\begin{figure}[!t]
  \centering
  \includegraphics[width=3.2in]{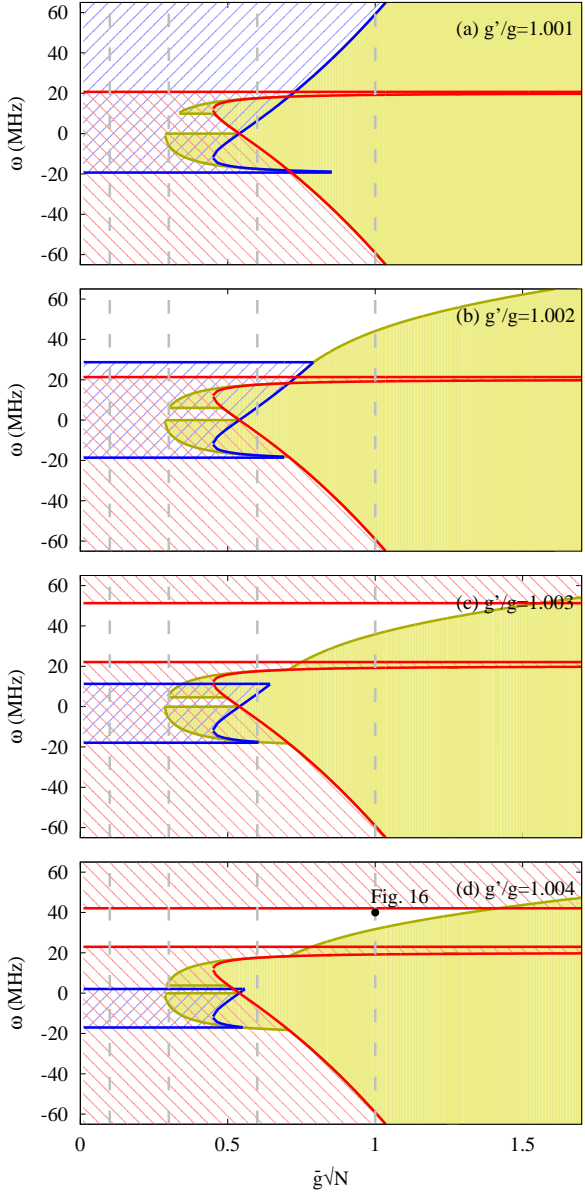}
  \caption{(color online).  
Evolution of the phase diagram shown in
    the bottom panel of Fig.~\ref{fig:omega-g-negative-u} as one
    varies the ratio $g^\prime/g$, increasing from 1.000 (top) to
    1.004 (bottom). We use the same phase labelling conventions as in
    Fig.~\ref{fig:omega-g-negative-u}. Vertical dashed lines indicate 
    cuts shown in Fig.~\ref{fig:omega-g-negative-u-vs-dg}.
}
  \label{fig:omega-g-negative-u-varying-ratio}
\end{figure}

\begin{figure}[!t]
  \centering
  \includegraphics[width=3.2in]{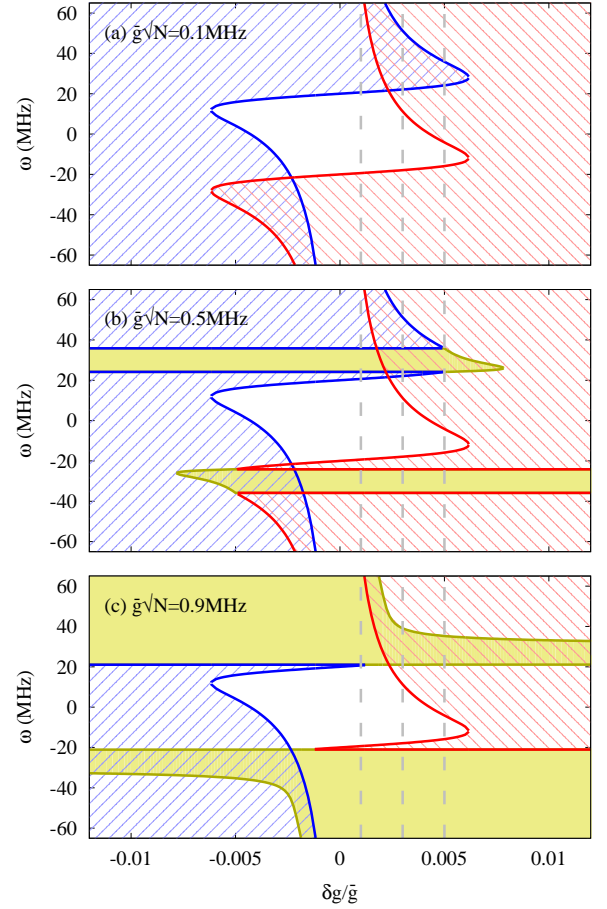}
  \caption{(color online).  
Analogue of the phase diagram in
    Fig.~\ref{fig:omega-g-negative-u-vs-dg} for $UN=+40$MHz. We use
    the same phase labelling conventions as in
    Fig.~\ref{fig:omega-g-negative-u}. The panels show the dependence
    on $\delta g\equiv g^\prime-g$ for fixed $\bar{g} = \frac{1}{2}
    (g+g^\prime)$. The white region corresponds to a regime of
    persistent oscillations. Vertical dashed lines indicate cuts shown
    in Fig.~\ref{fig:omega-g-positive-u-varying-ratio}.}
  \label{fig:omega-g-positive-u-vs-dg}
\end{figure}

\begin{figure}[!t]
  \centering
  \includegraphics[width=3.2in]{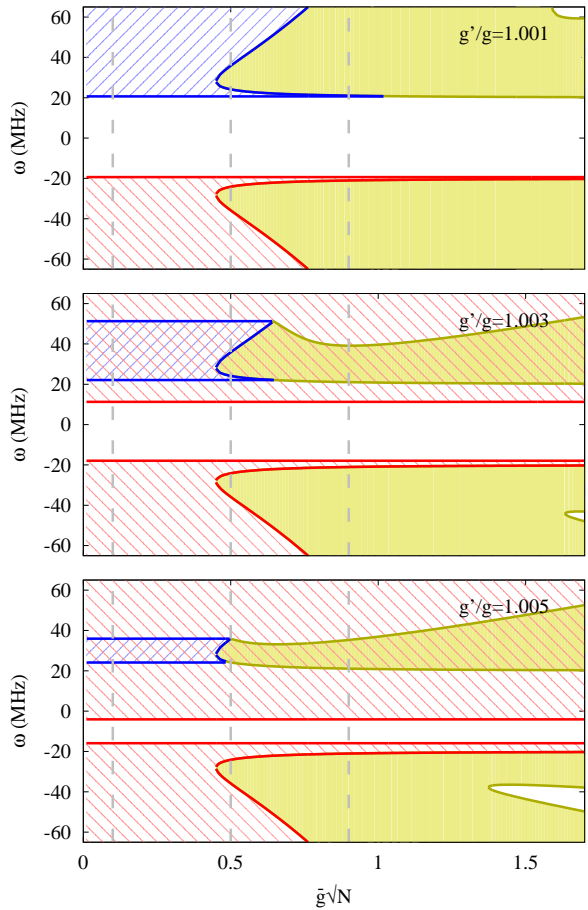}
  \caption{(color online).  
Analogue of the phase diagram shown in
    Fig.~\ref{fig:omega-g-negative-u-varying-ratio} for
    $UN=+40$MHz. We use the same phase labelling conventions as in
    Fig.~\ref{fig:omega-g-negative-u}. Vertical dashed lines indicate
    cuts shown in Fig.~\ref{fig:omega-g-positive-u-vs-dg}.}
  \label{fig:omega-g-positive-u-varying-ratio}
\end{figure}

In Sec.~\ref{sec:behaviour-g-neq} we presented the dynamical phase
diagram with $g\neq g^\prime$ and $UN=-40$MHz, illustrating the
dynamical phase boundaries which emerged for small differences between
$g$ and $g^\prime$. Here we provide further cuts through the phase
diagram in order to fully expose the rich topology.  In
Fig.~\ref{fig:omega-g-negative-u-varying-ratio}
we show a sequence of cuts with $UN=-40$MHz for different values of
$g^\prime/g$. These may be compared with the bottom panel of
Fig.~(\ref{fig:omega-g-negative-u}) which has $g^\prime/g=1$. In view
of the duality relation in Eq.~(\ref{duality}) we only show the
results for $g>g^\prime$.  For completeness, in
Figs.~\ref{fig:omega-g-positive-u-vs-dg}
and \ref{fig:omega-g-positive-u-varying-ratio} 
we also show cuts of constant $\bar{g}=(g+g^\prime)/2$ and
$g^\prime/g$ respectively, with $U=+40$MHz.  The central white region
in these figures is continuously connected to the regime of persistent
oscillations described in Sec.~\ref{sec:phase-diagram-u0}.  We note
that the regions where the normal and inverted states are stable have
identical shapes to those seen for $U=-40$MHz in
Fig.~\ref{fig:omega-g-negative-u-vs-dg} and in
Fig.~\ref{fig:omega-g-negative-u-varying-ratio} respectively, but are
displaced vertically.

\section{Wigner Distribution}
\label{App:Wigner}
Sampling initial conditions from a Wigner distribution is readily
achieved by combining a Holstein--Primakoff representation for the
collective spin operators, $S_z = -N/2 + a^\dagger a$ and $S^- \simeq
\sqrt{N} a$ \cite{Holstein:Primakoff,Auerbach:Interacting} with a
harmonic oscillator decomposition of the auxilliary bosons,
$a=(q+ip)/\sqrt{2}$. The corresponding Wigner distribution $W(q,p) =
e^{-q^2 - p^2}/\pi$ reflects the Gaussian ground state wavefunction of
the Harmonic oscillator.  In terms of the Bloch sphere coordinates
with $S^- = (N/2) \sin(\theta) e^{-i\phi}$ this corresponds to $\theta
\simeq \sqrt{(q^2 + p^2)/(N/2)}$ and $\phi=-{\rm Arg}(q+ip)$. In a
similar fashion, the Wigner distribution of the initial photon field
$\psi=(Q+iP)/\sqrt{2}$ is given by $W(Q,P) = e^{-Q^2 - P^2}/\pi$.  In
order to employ these ensembles we
sample initial conditions from these distributions of $(q,p)$ and
$(Q,P)$ and time evolve the semiclassical equations of motion. We then
average the final results over the initial distributions, $W(q,p)$ and
$W(Q,P)$.

\end{document}